\documentclass{emulateapj}
\usepackage{graphicx,graphics,amsmath}
\usepackage{natbib}
\usepackage{color}
\usepackage[normalem]{ulem}
\usepackage[plainpages=false, colorlinks=true, anchorcolor=blue,
linkcolor=blue, citecolor=blue, bookmarks=false]{hyperref}
\newcommand{\uvr}{\mbox{\boldmath $\hat{r}$}}
\newcommand{\uvt}{\mbox{\boldmath $\hat{\theta}$}}
\newcommand{\uvp}{\mbox{\boldmath $\hat{\phi}$}}
\newcommand{\del}{\mbox{\boldmath $\nabla$}}
\newcommand{\emf}{\mbox{\boldmath ${\cal E}$}}
\newcommand{\demf}{\mbox{\boldmath ${\cal D}$}}
\newcommand{\curl}{\mbox{\boldmath $\nabla \times$}}
\newcommand{\cross}{\mbox{\boldmath $\times$}}

\newcommand{\pd}{\partial}

\begin{document}

\title{Incorporating Surface Convection into a 3D Babcock-Leighton Solar Dynamo Model}
\author{Gopal Hazra$^{1}$, Mark S. Miesch$^{2}$}
\affil{$^1$Department of Physics, Indian Institute of Science, Bangalore 560012, India\\
$^2$National Oceanic and Atmospheric Administration, Boulder, CO 80305, USA}

\begin{abstract}
The observed convective flows on the photosphere (e.g., supergranulation, granulation) play a key role in the Babcock-Leighton (BL) process to generate large scale polar fields from sunspots fields. In most surface flux transport (SFT) and BL dynamo models, the dispersal and migration of surface fields is modeled as an effective turbulent diffusion.  Recent SFT models have incorporated explicit, realistic convective flows in order to improve the fidelity of convective transport but, to our knowledge, this has not yet been implemented in previous BL models.  Since most Flux-Transport (FT)/BL models are axisymmetric, they do not have the capacity to include such flows.  We present the first kinematic 3D FT/BL model to explicitly incorporate realistic convective flows based on solar observations.  Though we describe a means to generalize these flows to 3D, we find that the kinematic small-scale dynamo action they produce disrupts the operation of the cyclic dynamo.  Cyclic solution is found by limiting the convective flow to act only on the vertical radial component of the magnetic field. The results obtained are generally in good agreement with the observed surface flux evolution and with non-convective models that have a turbulent diffusivity on the order of $3 \times 10^{12}$ cm$^2$ s$^{-1}$ (300 km$^2$ s$^{-1}$).  However, we find that the use of a turbulent diffusivity underestimates the dynamo efficiency, producing weaker mean fields and shorter cycle than in the convective models.  Also, the convective models exhibit mixed polarity bands in the polar regions that have no counterpart in solar observations.  Also, the explicitly computed turbulent electromotive force (emf) bears little resemblance to a diffusive flux.  We also find that the poleward migration speed of poloidal flux is determined mainly by the meridional flow and the vertical diffusion.
\end{abstract}

\section{Introduction}\label{sec:intro}
Since it was discovered over a century ago that sunspots are regions of strong magnetic field \citep{Hale1909} and solar cycles are thus manifestations of solar magnetic activity, researchers have been seeking an explanation for this cyclic magnetic activity.  A major breakthrough in this history of research came with the work of \citet{parker55a} who first demonstrated how helical plasma motions can generate large-scale magnetic fields.  \citet{SKR66} later gave Parker's theory a more mathematical foundation with the development of mean field dynamo theory.  The first dynamo simulations to produce a solar-like butterfly diagram (cyclic equatorward migration of toroidal field) were made by \citet{SK69} and then \citet{Yoshimura75}. All of these calculations were based on the kinematic, mean field formulation of the magnetohydrodynamic (MHD) induction equation. 

Prior to the mid-1980s, when there was little observational information about the internal rotation profile of the Sun, theorists were free to adopt any profile that was needed in order to reproduce the observed butterfly diagram using the $\alpha$-$\Omega$ dynamo waves of mean-field theory.  However, by the mid 1990's the internal rotation profile of the Sun was well established \citep{thomp03}.  But, the use of this differential rotation profile in mean-field dynamo models was problematic because it adversely affected the propagation of $\alpha$-$\Omega$ dynamo waves.  \citet{CSD95} showed that the inclusion of a meridional circulation can help to solve this problem by advecting toroidal flux toward the equator, thus promoting more solar-like butterfly diagrams.  About the same time it was also argued that helical turbulence may not be as efficient at generating poloidal field as previously thought. This was based on non-kinematic effects associated with Lorentz-force back-reactions on both large and small scales. On large scales, it was argued that the toroidal field concentrations that give rise to sunspots may be too strong (super-equipartition fields of $\geq 10^4$ G) for convection to twist \citep{Dsilva93}. On small scales it was argued that the buildup of small-scale magnetic helicity may dramatically suppress the turbulent $\alpha$-effect \citep{vains92,gruzi94,brand01}. {Although, the turbulent dynamos including effects beyond those in simple $\alpha-\Omega$ models are capable of reproducing solar-like activity with helioseismic rotation profile and magnetic helicity conservation included but some of the issues in their model need to be investigated (e.g., stability of flux tube near surface layers against disruption due to magnetic buoyancy) \citep{Pipin13}.}

These challenges to traditional mean-field dynamo theory led to the resurgence of solar dynamo models based on the so-called Babcock-Leighton (BL) process.  The BL process (or BL mechanism) had been proposed decades earlier \citep{Bab61,Leighton69} as alternative candidate for the poloidal field generation that does not rely on turbulent convection.  In short, toroidal flux concentrations can destabilize and rise due to magnetic buoyancy, emerging from the photosphere as bipolar magnetic regions (BMRs).  The action of the Coriolis force on the rising flux tube induces a twist that is manifested upon emergence as a preferential tilt of BMRs, known as Joy's law, such that the trailing edge is displaced poleward relative to the leading edge.  The subsequent fragmentation and dispersal of these tilted BMRs after emergence due to turbulent diffusion, differential rotation, and meridional flow generates a dipole moment that, together with the $\Omega$-effect due to differential rotation, sustains the dynamo \citep{dikpa09,Charbonneau10,Karakreview14}. Note that the fragmentation and dispersal of the BMRs are executed by the convective flows on the photosphere (e.g., supergranulation and granulation) and this process is  modeled as a simple random of walk process which is treated as an effective turbulent diffusion \citep{Leighton64}. In the last two decades Babcock-Leighton (BL) solar dynamo models have grown to become the most promising paradigms to explain the origin of solar magnetic cycle and its irregularities \citep{CSD95,DC99,CNC04,dikpa09,Charbonneau10,Karak10,KarakChou11,Karakreview14}.   Yet, most of these models are still kinematic in the sense that they solve only the magnetohydrodynamic (MHD) induction equation with specified mean flows (differential rotation and meridional circulation) derived from observations.

Efforts to solve the fundamental MHD equations in the solar convection zone (CZ) from first principles in a more self-consistent manner date back to the pioneering work of \citet{GM81}, \citet{gilma83} and \citet{glatz84,glatz85a} and have made dramatic progress just in the last seven years.  Convective dynamo simulations now exhibit many solar-like features, including self-organization of large-scale fields, magnetic cycles with periods on the order of a decade, equatorward migration of toroidal flux, torsional oscillations, long-term cycle modulation, and even hints of magnetic flux emergence \citep{ghiza10,Racine11,Brown11,kapyl12,kapyl13,nelso13,nelso13b,passo14,fan14,warne14,Karak15,augus15,kapyl16,hotta16}. However, these models operate in parameter regimes far removed from the real solar interior and they still fall short of reproducing the solar cycle with high fidelity.  One reason may be that they do not yet have sufficient resolution to capture the full scope and complexity of the BL process.  This would require faithfully capturing the formation, rise, and emergence of the magnetic flux structures in the deep CZ, as well as the compressible, radiative MHD and small-scale convection in the surface layers responsible for the formation, fragmentation, and dispersal of active regions.

The amount of available flux emerging in BMRs each solar cycle, \citep[$> 2 \times 10^{24}$ Mx][]{schri94,thorn11} is two orders of magnitude larger than the amount of flux needed to reverse the dipole moment of the Sun, at least at the surface \citep[$\sim 5 \times 10^{22}$ Mx][]{munoz12}.  This alone suggests that the BL process may play an essential role in the solar dynamo.  This conclusion is further supported by the observed evolution of magnetic flux in the solar photosphere as represented by magnetic butterfly diagrams, which suggest that the polar field reversals are triggered by the poleward migration of magnetic flux originating from the trailing edge of BMRs \citep{Hathaway10a}.  Other lines of evidence in favor of the BL process include an observed correlation between the BL source term and cycle strength \citep{dasi10,Kitchatinov11a,mccli13}, the flux budget in active regions \citep{camer15}, and the phase relationship between poloidal and toroidal fields \citep{dikpa09,Charbonneau10,Karakreview14}.  

The observed evolution of magnetic flux in the solar photospheric is well captured by Surface Flux Transport (SFT) models. SFT models are not dynamo models. Magnetic flux is injected by means of artificial source terms or 2D magnetograms which provide the radial field $B_r$ as a function of latitude and longitude in some data assimilation window (typically the near side of the Sun).  The model then follows the subsequent evolution of this flux by solving a 2D (latitude-longitude) version of the kinematic MHD induction equation [eq.\ (\ref{induction1}) below] which includes differential rotation, meridional circulation, and turbulent diffusion by photospheric convection \citep{devor84,wang91}. As the evolution proceeds, the leading-polarity flux in low-latitude BMRs cancels across the equator while residual trailing-polarity flux is transported to the poles. The polarity of this trailing flux is opposite to the pre-existing polar flux so its accumulation at the poles from multiple BMRs eventually reverses the polar field and the global dipole moment.  This is the surface manifestation of the BL process.

\citet{upton14a,upton14b} have recently developed an SFT model called AFT (Advective Flux Transport) that used the observed surface flows to improve the fidelity of the model.  The distinguishing feature of AFT is that it uses explicit convective flow fields where other models use turbulent diffusion.  These 2D convective flow fields ($v_\theta$, $v_\phi$) are designed to reproduce the observed photospheric power spectrum and cell lifetimes, with randomized phases, as described by \citet{hatha00,Hathaway12a,Hathaway12}. \citet{upton14a,upton14b} showed that the use of these convective flows greatly improved the realism of the flux transport, reproducing the magnetic network and introducing stochastic variations that are not captured by a turbulent diffusion.

In this paper we use explicit three-dimensional (3D) convective flow fields for the first time in a Babcock-Leighton dynamo model of the solar cycle.  On the surface, these flow fields are identical to the empirical flow fields used in the AFT SFT model \citep{Hathaway12,upton14a,upton14b}.  However, here we extrapolate these flows below the surface to create a 3D rendition of surface convection that is responsible for the magnetic flux transport in the upper CZ.  Furthermore, in this initial implementation, we use a time-independent snapshot of the convective flow instead of the evolving flow fields used by Upton \& Hathaway.  We will implement evolving convective and mean flows in future work.

We achieve this through the use of the Surface flux Transport And Babcock-LEighton (STABLE) solar dynamo model \citep{MD14,MT16,HCM17}.  STABLE is a 3D model that explicitly places BMRs on the surface in response to the dynamo-generated toroidal field near the base of the CZ so the BL process can operate more realistically than in previous 2D (axisymmetric) models that employ, for example, a non-local $\alpha$-effect.  STABLE can function both as a BL dynamo model and as an SFT model, although we have not yet assimilated observational data in the latter context.  It is part of a next generation of 3D solar dynamo models that seek to capture the operation of the solar cycle with high fidelity by incorporating observational data and insights wherever possible.  In the future we will include Lorenz-force feedbacks by solving the full MHD equations but here we take the pragmatic approach of previous 2D models and solve the kinematic MHD induction equation with specified, realistic flow fields.  Since STABLE is 3D, these specified flow fields can include surface convection as well as differential rotation and meridional circulation.  Note that no existing global MHD convection simulation has sufficient resolution and scope to realistically capture surface convection on the scales of granulation to supergranulation.  So, our approach of specifying the convective velocity spectrum based on photospheric observations ensures that the surface convective motions are represented as realistically as is currently feasible.

Turbulent transport plays an important role in BL dynamo models, helping to regulate the cycle period and amplitude. 2D models typically represent it by a turbulent diffusivity and (sometimes) magnetic pumping but there are few observational constraints on how these coefficients vary with depth so models vary widely in the profiles they use \citep[see, e.g.][]{munoz11}.  Even SFT models vary in the coefficient they use to describe the observed dispersal of magnetic flux on the solar surface \citep{Jiang_review15}.   In this paper we investigate how realistic surface convection influences the behavior of a BL dynamo model.  Furthermore, by comparing convective models with models based on turbulent diffusion, we assess the viability of the turbulent diffusion paradigm and estimate the effective diffusion coefficient, $\eta_t$. Our results suggest that the surface convection is approximately equivalent to a turbulent diffusion: $\eta_t \sim 3 \times 10^{12}$ cm$^2$ s$^{-1}$ which is comparable with the values obtained from observations: $\eta_t \sim$ 2--5 $\times 10^{12}$ cm$^2$ s$^{-1}$ \citep{mosher77,topka82,Schrijver96,chae08,Jiang_review15}

The organization of the paper is as follows. In Section~2 and 3, we describe the STABLE model and how we implement convective flow fields.  In Sections~4 and 5 we describe dynamo simulations based on turbulent diffusion and explicit convective motions respectively.  In Section~6 we compare and contrast these simulations in order to quantify the role of convective transport and estimate the effective turbulent diffusivity of the convective motions.  We conclude and summarize our results in Section~7.
       
\section{The Numerical Model}\label{sec:STABLE}
The Surface flux Transport And Babcock-LEighton (STABLE) dynamo model is a 3D generalization of previous 2D BL models.  In particular, it can be classified as a Flux-Transport Dynamo (FTD) model \citep{WSN91,CSD95,DC99,dikpa09}.  FTD models are BL dynamo models in which the imposed meridional circulation plays an important role in transporting toroidal flux toward the equator, thus establishing the butterfly diagram.  Toroidal field is generated by the stretching of poloidal field by the differential rotation (the $\Omega$-effect) and the poloidal field is generated by the BL process (Sec.\ \ref{sec:intro}).  

Many previous 2D models parameterize the BL process by means of an $\alpha$-effect or similar poloidal source term that is nonlocal in the sense that the poloidal field generation is confined to the surface layers but depends on the toroidal flux near the base of the CZ \cite[e.g.][]{DC99,rempe06}.  We take a different approach here, exploiting the 3D capabilities of STABLE to treat the inherently 3D BL process more realistically.  In particular, we explicitly place tilted BMRs on the surface of the model and follow their subsequent evolution by solving the 3D kinematic MHD induction equation:
\begin{equation}\label{induction1}
\frac{\partial {\bf B}}{\partial t} = \nabla \times ({\bf v}\times{\bf B} -\eta_t \nabla\times {\bf B}) ~~~.
\end{equation} 
The BMRs are sheared out and dispersed by differential rotation and turbulent diffusion and advected poleward by the differential rotation, naturally generating mean poloidal field as originally described by \citet{Bab61} and \citet{Leighton69} and as captured by SFT models (Sec.\ \ref{sec:intro}).  

The numerical algorithm that places BMRs on the surface is called SpotMaker and is described in detail by \citet{MD14} and \citet[][hereafter MT16]{MT16}.  Briefly, we define a spot-producing toroidal field $B^*(\theta,\phi,t)$ at any time $t$ by integrating the longitudinal field component $B_\phi(\theta,\phi,r,t)$ over a radial range near the base of the CZ (here spanning 0.7--0.71$R$, where $R$ is the solar radius) and applying a latitudinal mask that suppresses high latitudes.  We then choose a random latitude and longitude from all points where $B^*(\theta,\phi,t)$ exceeds a threshold value (here equal to 1 kG).  This is where we place a tilted BMR on the surface.

The tilt angle of each BMR follows Joy's law ($\delta = 32^\circ.1 \cos\theta$; Stenflo \& Kosovichev 2012\nocite{SK12})  and the subsurface spot structure is specified by means of a potential field extrapolation down to $r_p = 0.90 R$.  As discussed by MT16 \citep[see also][]{Longcope02,SM05}, this corresponds to the limit in which BMRs decouple quickly from their roots in the tachocline/lower CZ (on a time scale short compared with the solar cycle).  The subsurface structure is a very idealized assumption but it serves to localize the BL process in the near-surface layers as in previous FTD models.  In the future we will consider the opposite limit in which the BMR retains connectivity to the tachocline by implementing a lifting and twisting flow in SpotMaker as described by \citet{YM13}.  

In its current rendition, SpotMaker is essentially a 3D generalization of Durney's Double ring algorithm \citep{Durney95,Durney97,Munoz10}.  It effectively serves as an explicit source term $S(r,\theta,\phi,t)$ added to equation (\ref{induction1}): 
\begin{equation}
\label{induction2}
\frac{\partial {\bf B}}{\partial t} = \nabla \times ({\bf v}\times{\bf B} -\eta_t \nabla\times {\bf B}) + S(r,\theta,\phi, t)  ~~~.
\end{equation}  
Expressed in this way, $S(r,\theta,\phi,t)$ would be composed of a series of $\delta$ functions in time, with several thousand instances in each 11-year cycle.  The time interval between spot appearances in each hemisphere is chosen randomly from a lognormal probability distribution with a mean of 3 days and a mode of 2 days (see Fig.\ 2\textit{c} in MT16).  

The magnetic flux in each BMR is proportional to the toroidal flux at the base of the CZ, $B^*(\theta,\phi,t)$, and at the chosen location, $\theta_s$, $\phi_s$:
\begin{equation}
\label{flux}
\Phi = 2\alpha_{spot}\frac{|{\hat{B}(\theta_s,\phi_s,t)}|}{B_q}\frac{10^{23}}{1 + (\hat{B}(\theta,\phi)/B_q)^2} \rm{Mx}
\end{equation}  
We neglect the rise time of a flux tube, which is short ($\sim$ 1 month) compared to a cycle period ($\sim$ 11 years).  We also currently neglect the deflection of the tube as it rises.  So, the flux in each BMR is proportional to the instantaneous value of $\hat{B}$ at the same time, latitude, and longitude, as described by MT16.  For the simulations reported here we use a quenching field strength $B_q$ of 100 kG.

Defining the flux content of the spots in this way (equation \ref{flux}) has various advantages. The poloidal field generation is proportional to $\hat{B}$ for weak fields, so it correctly mimics the Babcock-Leighton $\alpha$ effect. For $\alpha_{spot}$ = 1, the subsurface field at the quenching strength $B_q$ will generate the biggest spots that are observed which have a flux content $\sim 10^{23} \rm{Mx}$.  However, as discussed by MT16, it is sometimes necessary to artificially increase the BMR flux in order to achieve supercritical (non-decaying) dynamo solutions.  This is done by choosing a  value for $\alpha_{spot} > 1$.  Hence, this parameter determines how much poloidal field will be generated from the subsurface toroidal field and helps to regulate the overall amplitude of the dynamo fields.  We define the radius of a spot $r_r$ based on its flux content, such that $\Phi = B_0 r_r^2$ \citep[neglecting geometric factors of order unity; see][]{MD14}.  For the value of $B_0$ we choose a typical sunspot field strength of 3 kG.  We impose a minimum value of $r_r$ based on the spatial resolution. 

We use the well-tested Anelastic Spherical Harmonic (ASH) code \citep{Miesch_et_al_00,BMT04} to solve only the 3D magnetic induction equation (\ref{induction2}), with fixed velocity fields.  This kinematic formulation follows the pragmatic approach of previous 2D FTD dynamo models that take advantage of solar observations and helioseismic inversions to make the flow fields as realistic as possible.  We will consider Lorentz-force feedbacks in future work.  The ASH code has been verified against both convective dynamo benchmarks \citep{jones11} and axisymmetric FTD benchmarks (MT16).   

The velocity field ${\bf v}$ in eq.\ (\ref{induction2}) includes differential rotation (DR) and meridional circulation (MC) profiles chosen to match helioseismic inversions and photospheric observations, where available.  Despite a recent proliferation of research fueled by the availability of high-resolution HMI/SDO data and long time series from MDI/SOHO, the subsurface structure and amplitude of the MC is still uncertain.  There is some possible evidence for multiple cells per hemisphere in latitude and radius but different methods do not yet provide a consistent picture \citep{Hathaway12,Zhao13,Schad13,RA15,jacki15}.  In light of this uncertainty and in order to be consistent with the vast majority of papers in the literature on FTD dynamos, we adopt an MC profile here with a single cell per hemisphere, with counter-clockwise circulation in the northern hemisphere (NH) and clockwise circulation in the southern hemisphere (SH).  This yields a peak poleward flow of 25.5 m s$^{-1}$ at mid-latitudes at the surface and an equatorward flow of $\sim $ 2 m s$^{-1}$ near the base of the CZ.  For further details on the MC profile we use see \citet{CNC04} and \citet[][hereafter HCM17]{HCM17}.  

The DR profile is chosen to be independent of radius in the bulk of the CZ, with a latitudinal rotation rate varying from 460.7 nHz at the equator to 330.67 nHz at the poles.  We do not include a near-surface shear layer but we do include a tachocline, expressed by means of an error function centered at $r_c = 0.7 R$, with a thickness of 0.05 $R$.  This produces a sharp transition at the base of the CZ to uniform rotation rate of the radiative zone, which is set equal to 432.8 nHz.  These values are based on the previous 2D FTD models of \citet{DC99} and \citet{CNC04} and our implementation is described further by MT16 (see Fig.\ 1$a$).  \citet{dikpati02} have argued that the near-surface shear layer does not contribute significantly to the operation of flux-transport solar dynamo, but \citet{karak16} have shown that it can promote equatorward propagation in conjunction with efficient magnetic pumping in the surface layers. 

\begin{figure}[!h]
\centering
\includegraphics[width=7.5 cm,height = 7 cm, angle=0]{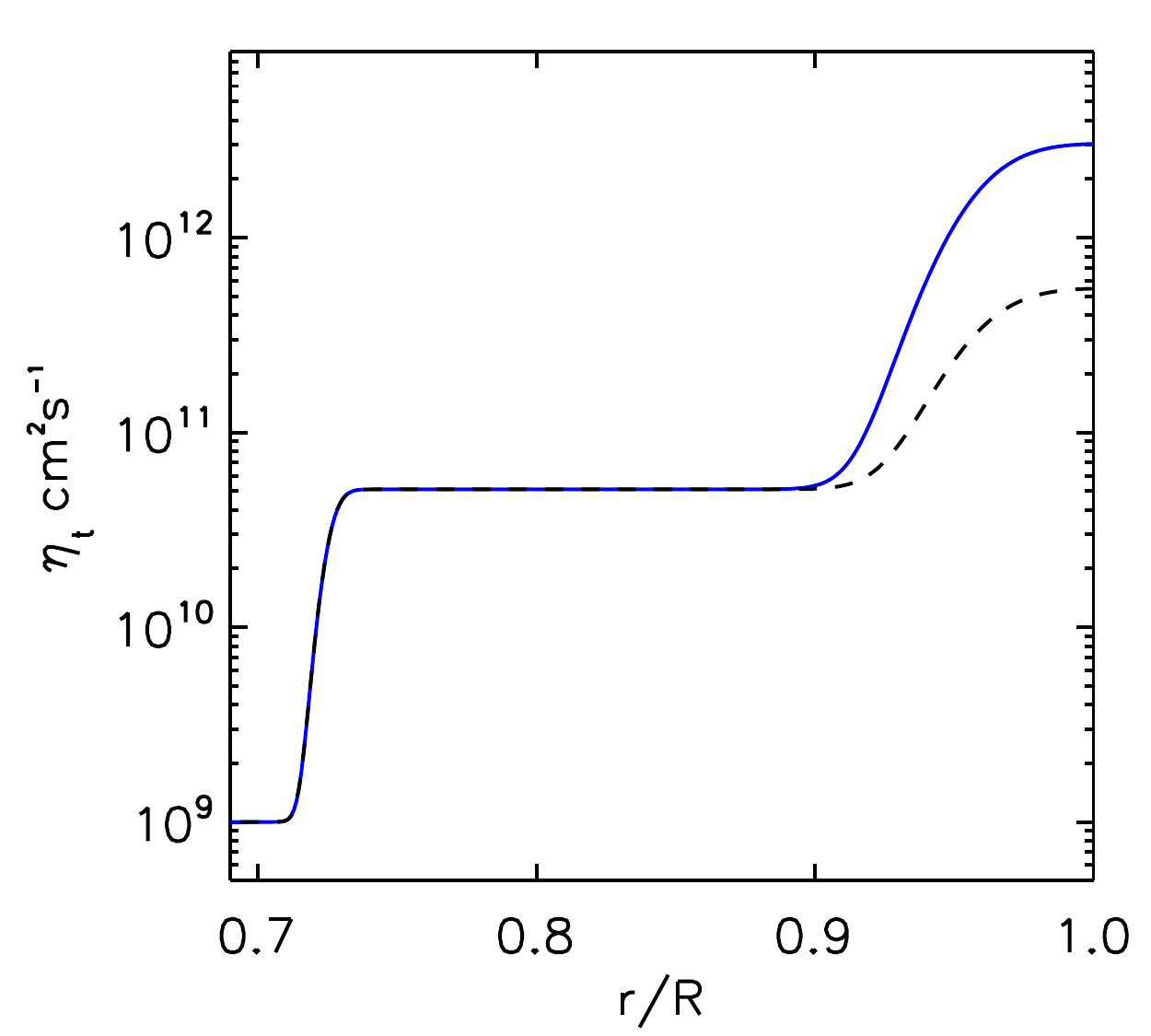}
\caption{Two step diffusivity profile used in our simulations. Shown are example diffusivity profiles used for simulations with convection (black dotted line) and without explicit convective motions (solid blue line), with $\eta_{top} = 3\times 10^{12}$ cm s$^{-1}$ in the latter.  The convective motions are confined to the upper convection zone and their role in the global dynamo is functionally equivalent to an enhanced turbulent diffusion.}
\label{fig:eta}
\end{figure}

In this paper for the first time, we also include a non-axisymmetric convective velocity component in ${\bf v}$.  Our methodology for achieving this is described in Sec.\ \ref{sec:cflow}.  This convective flow is confined to the upper CZ ($r > 0.9R$) and is intended to capture magnetic flux transport by small-scale surface convection on scales from granulation to super-granulation.  Transport in the deeper regions of the CZ (giant cells) is still handled by means of the turbulent diffusivity $\eta_t$, which is a function of radius alone.  For the simulations described in Section \ref{sec:nocon} that do not include 3D convective motions we use a two-step diffusivity profile as in previous 2D FTD models \citep[e.g.][]{Hotta10}.  Example profiles are shown in Fig.\ \ref{fig:eta} (see also Fig.\ 1\textit{c} in MT16 and Fig.\ 3 in HCM17).  The value of $\eta_t(r)$ at $r=R$, which we refer to as $\eta_{top}$, is varied between 1$\times 10^{10}$ cm$^2$ s$^{-1}$ and $3.5 \times 10^{13}$ cm$^2$ s$^{-1}$, as described in Sec.\ \ref{sec:nocon}.  The values of $\eta_t(r)$ in the lower CZ and radiative interior are respectively $5\times 10^{10}$ cm$^2$ s$^{-1}$, and 10$^9$ cm$^2$ s$^{-1}$.  For the simulations described in Sec.\ \ref{sec:con}, we reduce the surface diffusion to ensure that the transport in the upper CZ is dominated by the imposed convective motions.

Note that the value of $\eta_t$ that we use in the mid CZ, $5 \times 10^{10}$ cm$^2$ s$^{-1}$ puts us in the so-called advection-dominated regime of FTD models in the sense that the MC dominates the transport of poloidal magnetic flux from the surface layers to the base of the CZ \citep{Jiang07,Yeates08,Karakreview14}.  STABLE can also operate in the diffusion-dominated regime in which turbulent diffusion is the dominant transport mechanism; for an example see HCM17.  It has been argued on several different grounds that the diffusion-dominated regime may be more realistic \citep[e.g.][]{CCJ07,Jiang07,KarakChou11,miesc12} and is less sensitive to the detailed structure of the MC, capable of producing solar-like cycles even for multi-cellular MC profiles \citep{HKC14}.  However, an investigation of these issues lies outside the scope of this paper.  Here we focus on how explicit convective transport affects the operation of FTD dynamo models and our chosen $\eta_t(r)$ profiles are sufficient for this purpose.

We also include a weak horizontal hyperdiffusion on the right-hand-side of eq.\ (\ref{induction2}) to dissipate spurious small-scale fields and thus keep the code numerically stable.  In spectral space this is expressed as $- \eta_h R^{-2} [\ell (\ell+1)]^2 \widetilde{\bf B}$, where $\widetilde{\bf B}$ is the spherical harmonic transform of ${\bf B}$.  We use $\eta_h = 2 \times 10^8$ cm$^2$ s$^{-1}$ for the diffusive Cases A1-A8 (Sec.\ \ref{sec:nocon}) and $\eta_h = 2\times 10^{10}$ cm$^2$ s$^{-1}$ for the convective Cases C1-C3 (Sec.\ \ref{sec:con}).

\begin{figure*}[!t]
\centering
\includegraphics[width=\textwidth]{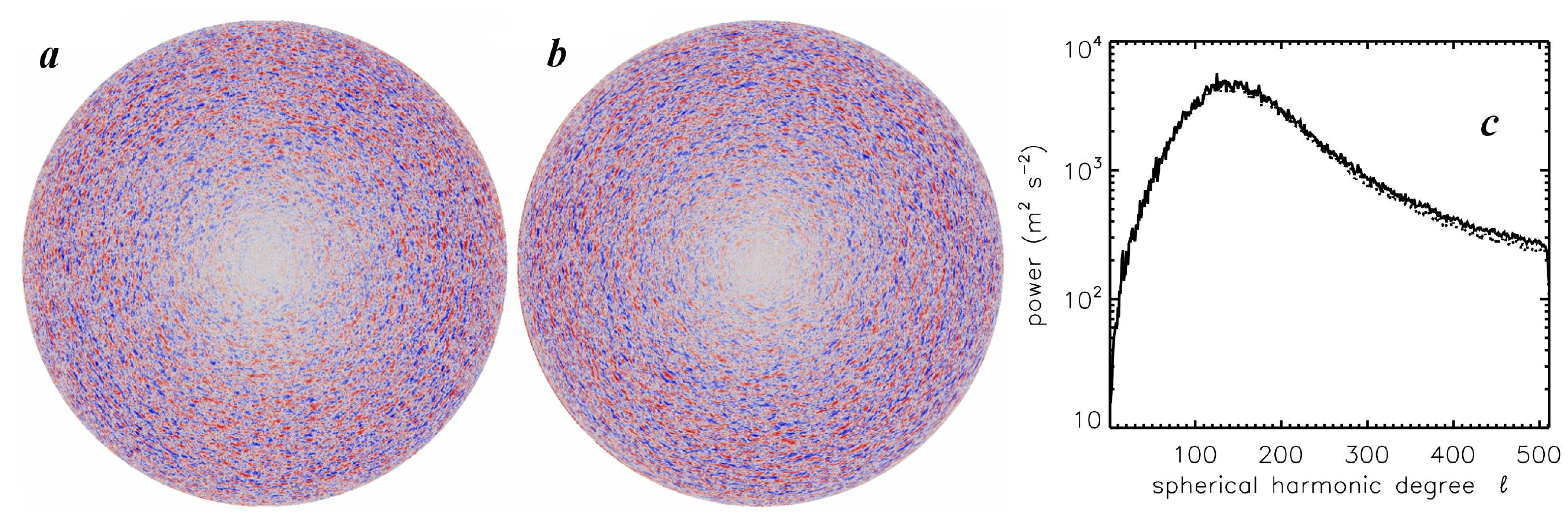}
\caption{(\textit{a}) Line-of-sight (Doppler) velocity in the solar photosphere measured with the SOHO/MDI instrument on June 4, 1996. (\textit{b}) Simulated line-of sight velocity constructed from the empirical model of \citet{hatha00} and \citet{Hathaway12a,Hathaway12}, which is designed to reproduce the observed horizontal Doppler velocity power spectrum, with randomized phases.  (\textit{c}) Horizontal velocity spectra for Hathaway's simulated flow field (solid line) and for the convective flow field used here (dotted line), at $r=R$.  Small discrepancies at large $\ell$ arise because we only extract the divergent component (see text).  Images and data courtesy of David Hathaway and Lisa Upton.}
\label{fig:conhath}
\end{figure*}

\section{Incorporating Non-Axisymmetric Convective Flows into STABLE}\label{sec:cflow}

\subsection{Empirical Model for Surface Convection}\label{sec:emp}
Though ASH is fully capable of simulating global-scale convective motions, no global solar convection model can accurately capture smaller-scale convective motions near the surface such as granulation and supergranulation.  This would require both extremely high resolution and including physical processes that are often neglected in global models, such as non-LTE radiative transfer, ionization, and the breakdown of the anelastic approximation.  However, it is these small-scale convective motions that contribute most to the breakup and dispersal of BMRs and are thus most important from the perspective of the BL process.

For this reason, we wish to incorporate convection in a manner that is consistent with the pragmatic approach of kinematic dynamo modeling.  In particular, we wish to exploit observational measurements of the convective power spectrum in the solar photosphere in order to ensure that the imposed convective flow fields near the surface are as realistic as possible.  High-quality, full-disk measurements of the photospheric convection spectrum are now available from such instruments as the Helioseismic Magnetic Imager (HMI) onboard NASA's Solar Dynamics Observatory and the Michelson Doppler Imager (MDI) onboard NASA's SOlar and Heliospheric Observatory (SOHO).   These measurements are typically based on Dopplergrams; 2D (latitude-longitude) maps of the line-of-sight velocity component on the solar surface (Fig.\ \ref{fig:conhath}\textit{a}).

Since the velocities in the solar photosphere are predominantly horizontal, these are the velocity components that are mainly sampled by the Dopplergrams; this leads to the dearth of power at disc center in Fig.\ \ref{fig:conhath}\textit{a}.  Furthermore, at any point on the solar disc, only one horizontal direction is sampled.  For example, Doppler velocities near the eastern limb are dominated by $V_\phi$.  In order to reconstruct the complete 2D flow field $V_\theta(\theta,\phi,t)$ and $V_\phi(\theta,\phi,t)$, some modeling is needed. \citet{Hathaway12a,Hathaway12} has devised such a model \citep[see also][]{hatha00}.  Hathaway's model is based on first subtracting off the differential rotation, the meridional circulation, and other unwanted signals such as convective blueshift, spacecraft motion and instrumental artifacts.  Then the convective power spectrum is computed.   A simulated horizontal velocity field is then produced based on that observed spectrum using a series of vector spherical harmonics with randomized complex coefficients.  

Though Hathaway's original implementation is time-evolving, with correlation times chosen to match observations, we consider here a static horizontal velocity field with a resolution of $N_\theta = 512$ and $N_\phi = 1024$.  A sample Dopplergram computed from the simulated horizontal flow field is shown in Fig.\ \ref{fig:conhath}\textit{b} and the horizontal power spectrum is shown in Fig.\ \ref{fig:conhath}\textit{c}.  The peak at $\ell \approx $ 130 represents supergranulation.  This data set does not resolve the broad peak in power beyond $\ell \sim 1000$ due to granulation, as discussed by \citet{hatha00}.  We will refer to this empirical surface velocity field as ${\bf V}_s(\theta,\phi,t)$.  Though we consider only static flows fields in this paper, we will retain the explicit time dependence in this section in order to illustrate how our approach can be readily generalized to evolving convective flows.

Our aim is to incorporate the empirical surface velocity field ${\bf V}_s(\theta,\phi,t)$ into STABLE.  Since this velocities are non-axisymmetric, this exploits the 3D capabilities of our dynamo model.  In particular, this implies that the transport and amplification of the non-axisymmetric magnetic field components can influence the time evolution of the mean fields.  This is not the case for axisymmetric, kinematic flows fields in which the $m=0$ component of the magnetic induction equation decouples from the $m>0$ components, where $m$ is the azimuthal wavenumber.

In order to incorporate the 2D empirical surface flow field of Fig.\ \ref{fig:conhath} into our 3D model, we must specify some radial structure.  In specifying this radial structure we have made several assumptions and approximations.  The first assumption is that the small-scale convection in the solar surface layers is much more vigorous than the convection in the deeper convection zone.  This is justified by our theoretical understanding of solar convection, which attributes it to the relatively low density, steep density stratification, and steep superadiabatic entropy gradient in the solar surface layers \citep{miesc05,nordl09}.  It is also justified by recent attempts to estimate the convective velocity amplitudes in the deep solar interior by means of helioseismic inversions, photospheric observations, and numerical modeling \citep{hanas10,miesc12,lord14,greer15,hanas16,omara16,cosse16}.  And, it is consistent with previous FTD models that often use an enhanced turbulent diffusivity in the solar surface layers as illustrated in Fig.\ \ref{fig:eta}.  Thus, we wish to extrapolate the surface velocity field downward but no deeper than $r \sim 0.9$, so that it effectively replaces the enhanced turbulent diffusion near the surface as shown by the dashed line in Fig.\ \ref{fig:eta}.

The second assumption is that the mass flux is divergenceless: $\nabla \cdot (\overline{\rho} {\bf V}) = 0$.  Here $\overline{\rho} = \overline{\rho}(r)$ is the background density stratification, neglecting density fluctuations associated with the convection.  Thus, this is essentially an anelastic approximation, valid for low Mach numbers.  This means that the mass flux can be decomposed into poloidal $(W)$ and toroidal $(Z)$ components defined by:
\begin{equation}\label{eq:poltor}
\overline{\rho} {\bf V} = \nabla \times \nabla \times \left(W \uvr\right) + \nabla \times \left(Z\uvr\right)  ~~~.
\end{equation}

\begin{figure}[!h]
\centering
\includegraphics[width=0.5\textwidth]{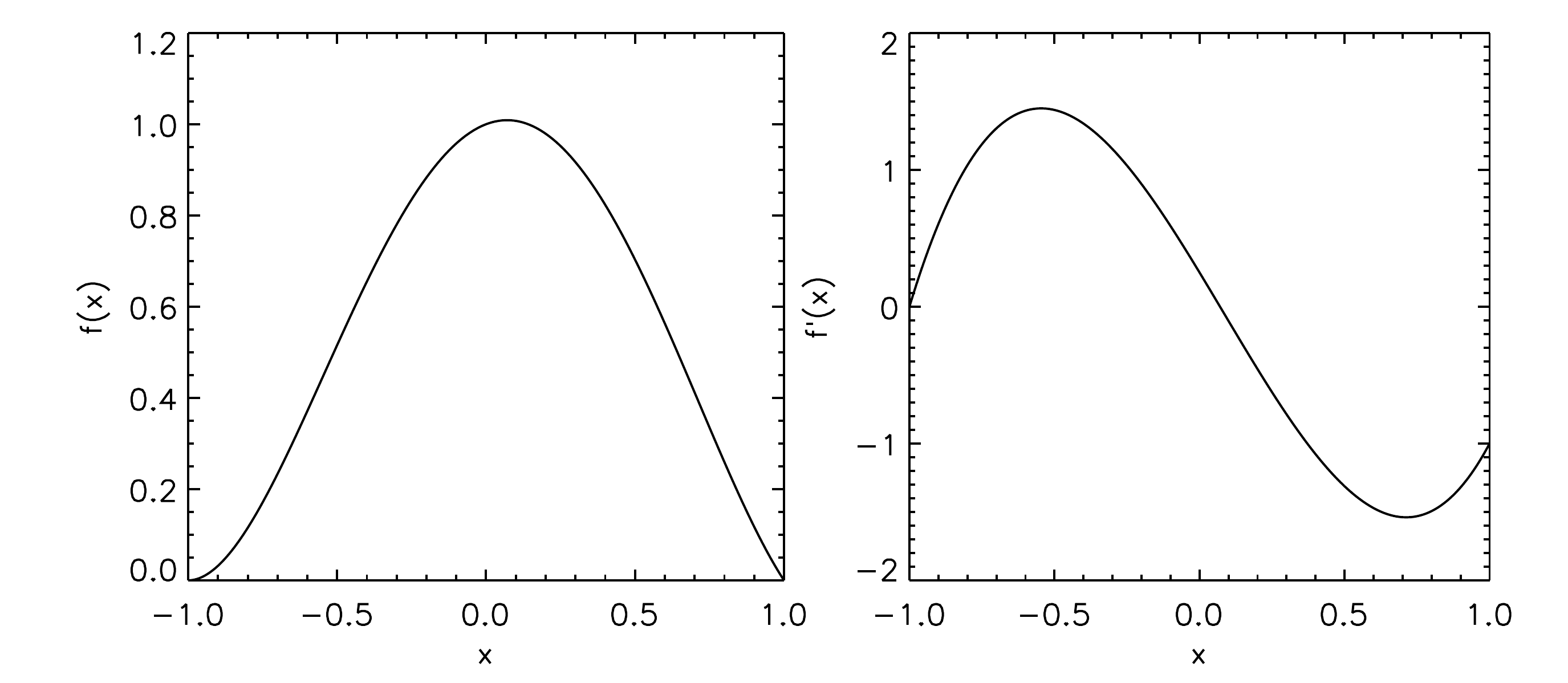}
\caption{Radial polynomial used for extrapolation of the surface flow to the deeper convection zone. Shown are (\textit{a}) $f_\ell(x)$ and (\textit{b}) $f_\ell^\prime$(x), where $x = (r-r_w)/(R-r_w)$. The plots extend from $r=r_p$ ($x=-1$) to $r=R$ ($x=1$).}
\label{fig:radfnc}
\end{figure}

Our third assumption is that the convective velocities are purely poloidal, so $Z = 0$.  This implies that the vertical component of the fluid vorticity is zero.  This is justified mainly by the steep density stratification in the solar surface layers, which imparts a strong horizontal divergence to vertical motions, and the clear signature of supergranulation in the horizontal divergence of surface flows \citep{gizon10}.  There is also clear evidence that the action of the Coriolis force imparts a systematic vertical vorticity to supergranular-scale convective motions \citep{gizon10} but this is a relatively weak effect and it is justified to neglect it as a first approximation.  In future work we will include a vortical (toroidal) component to the flow field and we will investigate its influence both on the turbulent transport and on the potential amplification of magnetic flux.  

This in effect means that we only assimilate the divergent component of the surface flow field into STABLE.  Thus, we define $D(\theta,\phi,t)$ as the horizontal divergence of the imposed surface flow field: $D(\theta,\phi,t) = \del_h \cdot {\bf V}_s$.  Recall that ${\bf V}_s$ has only horizontal components so $D$ is also equal to the full divergence of the surface flow field.  Now we relate the horizontal divergence of the full 3D convective flow field to the poloidal stream function $W$ as follows:
\begin{equation}\label{vh}
\rho\del_h \cdot {\bf V}_h = -\frac{1}{r^2}\frac{\partial}{\partial r}(r^2\rho V_r) 
= - \nabla_h^2 \left(\frac{\partial W}{\partial r}\right) 
\end{equation}
Where $\nabla_h^2$ and ${\bf V}_h = V_\theta \uvt + V_\phi \uvp$ are the horizontal Laplacian and the horizontal velocity respectively.

Now we expand $W$ in a spherical harmonic series as follows
\begin{equation}\label{wspectral}
W(r,\theta,\phi,t) = \sum_{\ell=0}^{\ell_{max}} \sum_{m=-\ell}^\ell \tilde{g}_{\ell m}(t) f_\ell(r) Y_{\ell m}(\theta,\phi)  ~~~.
\end{equation}
Here $\tilde{g}_{\ell m}(t)$ describes the horizontal structure of the convective pattern at the surface and $f_\ell(r)$ describes its downward extrapolation.  We have included an explicit $\ell$ dependence for $f_\ell(r)$ to allow for the possibility that different horizontal scales of convection may have different radial profiles; see Sec.\ \ref{sec:fr}.  Note also that
\begin{equation}\label{eq:hlap}
- \nabla_h^2 W = \sum_{\ell=0}^{\ell_{max}} \sum_{m=-\ell}^\ell \frac{\ell (\ell+1)}{r^2} ~ \tilde{g}_{\ell m} f_\ell Y_{\ell m}  ~~~.
\end{equation}

An expression for $\tilde{g}_{\ell m}$ can be obtained by applying a spherical harmonic transform to eq.\ (\ref{vh}) and evaluating it at the surface, $r=R$, with the help of eq.\ (\ref{eq:hlap}).  This yields:
\begin{equation}\label{gtheta}
\tilde{g}_{\ell m}(t) = \frac{R^2 \overline{\rho}(R)}{\ell(\ell+1)} ~ \tilde{D}_{\ell m}(t) ~ \left[f_\ell^\prime(R)\right]^{-1}  ~~~,
\end{equation}
where $\tilde{D}_{\ell m}(t)$ are the spherical harmonic coefficients for the surface divergence $D(\theta,\phi,t)$ and $f_\ell^\prime(r) = df_\ell(r)/dr$. All that remains is to define the vertical profile $f_\ell(r)$, which we discuss in Section \ref{sec:fr}.  

Once we define $\tilde{g}_{\ell m}(t)$, $f_\ell(r)$, and $\overline{\rho}(r)$, then we can obtain all three components of the convective flow field throughout the entire 3D computational domain by means of equations (\ref{wspectral}) and (\ref{eq:poltor}):
\begin{eqnarray}
V_r(r,\theta,\phi,t) &=& - \frac{1}{\overline{\rho}} \nabla_h^2 W(r,\theta,\phi,t) \\
V_\theta(r,\theta,\phi,t) &=& \frac{1}{\overline{\rho} r}\frac{\partial}{\partial \theta} \left(\frac{\partial W}{\partial r}\right) \\
V_\phi(r,\theta,\phi,t) &=& \frac{1}{\overline{\rho} r \sin\theta}\frac{\partial}{\partial \phi} \left(\frac{\partial W}{\partial r}\right)
\end{eqnarray}
Again, in this paper we consider a time-independent convective flow field but this is easily generalizable to evolving flows in which the time evolution is governed by the imposed, empirical surface flow field ${\bf V}_s(\theta,\phi,t)$.  This is in turn reflected in $\tilde{g}_{\ell m}(t)$ through eq.\ (\ref{gtheta}).

Equation (\ref{gtheta}) is the means by which Hathaway's simulated surface flow field, ${\bf V}_s$, is assimilated into STABLE.  As noted above, only the horizontal divergence is used so any non-divergent components of ${\bf V}_s$ will be omitted from our flow field.  Hathaway's horizontal flow field is also constructed on the assumption that the flow is strictly poloidal ($\curl {\bf V}_s = 0$) so this procedure should in principle ensure that $V_\theta$ and $V_\phi$ reproduce ${\bf V}_s$ exactly at $r=R$.  However, in practice there are numerical errors associated with the discretization of ${\bf V}_s$, the interpolation onto a Legendre grid for incorporation into STABLE, and the numerical computation of the derivatives (though the latter computation is spectrally accurate).  So, the resulting horizontal power spectrum lies slightly below Hathaway's spectrum at high wavenumbers, as shown in Fig.\ \ref{fig:conhath}\textit{c}.

For the background density we use a polytropic, hydrostatic, adiabatic stratification \citep{jones11}:
\begin{equation}\label{eq:rho}
\overline{\rho} = \rho_i ~ \left(\frac{\zeta(r)}{\zeta(r_1)}\right)^n
\end{equation}
where
\begin{equation}
\zeta(r) = c_0 + c_1 \frac{r_2-r_1}{r}
\end{equation}
\begin{equation}
c_0 = \frac{2 \zeta_0 - \beta - 1}{1-\beta}
\end{equation}
\begin{equation}
c_1 = \frac{(1+\beta)(1-\zeta_0)}{(1-\beta)^2}
\end{equation}
and
\begin{equation}
\zeta_0 = \frac{\beta+1}{\beta \exp(N_\rho/n) + 1}  ~~~.
\end{equation}
Here $\rho_i = \rho(r_1) = 0.1788$ g cm$^{-3}$, $\beta = r_1/r_2 = 0.69$ is the aspect ratio,
$n = 1.5$ is the polytropic index, and $N_\rho = 5$ is the number of density scale heights
across the computational domain, which extends from $r_1 = 0.69 R$ to $r_2 = R$.

\begin{figure*}[t!]
\centering
\includegraphics[width=\textwidth]{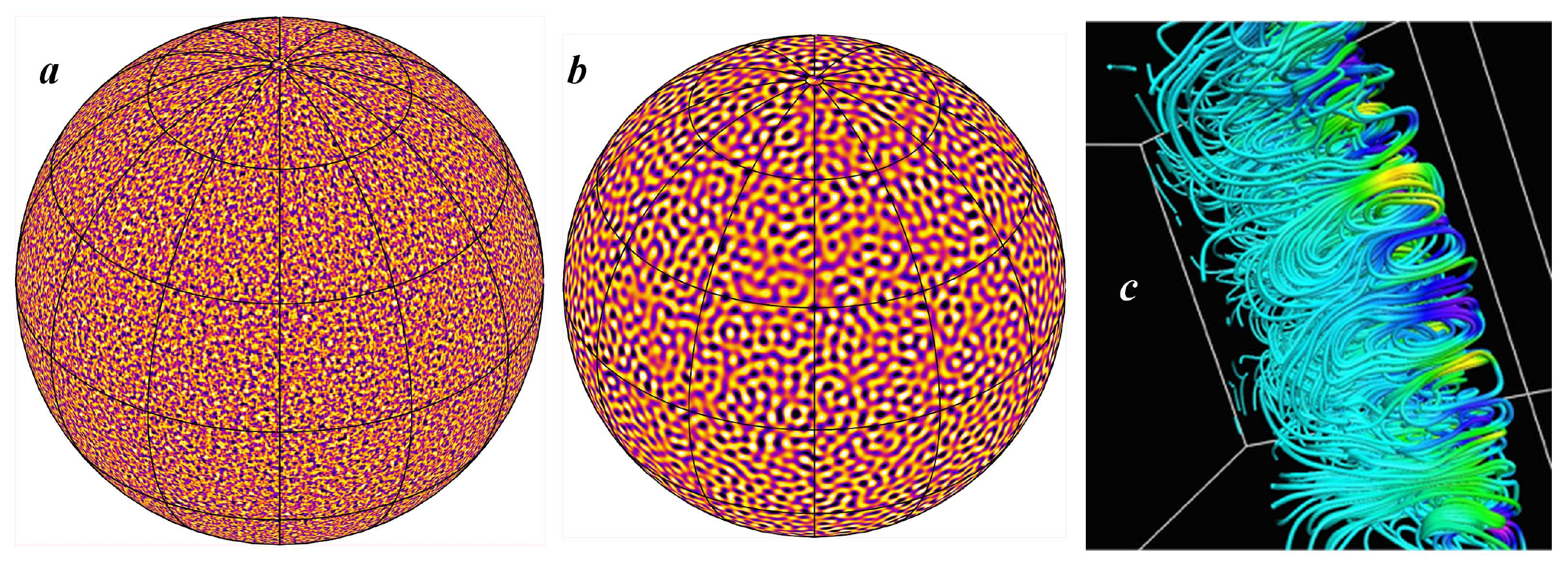}
\caption{Visualization of the imposed convective flow field.  (\textit{a}) The horizontal divergence at the surface $r=R$ is set equal to the divergence of the empirical flow field ${\bf V}_S(\theta,\phi)$.  Here it is shown in an orthographic projection with yellow and blue representing divergence and convergence respectively. (\textit{b}) As in (\textit{a}) but for $r=0.952$. (\textit{c}) 3D rendering of a zoomed-in portion of the flow field showing streamlines of the mass flux colored by the vertical velocity (yellow upward, blue downward).  Velocity amplitudes decrease with depth due to the increasing density.}
\label{fig:viz1}
\end{figure*}

\begin{figure*}[]
\centering
\includegraphics[width=\textwidth]{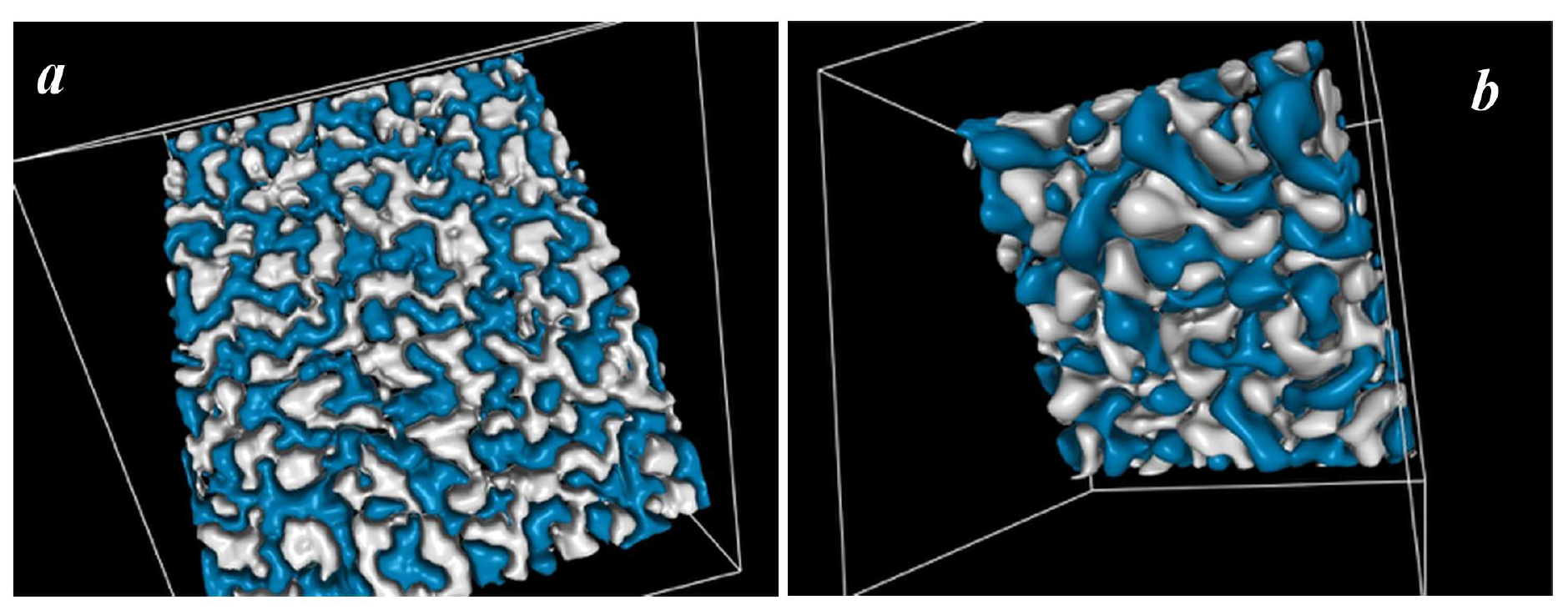}
\caption{Isosurfaces of the convective radial velocity $V_r$ for a zoomed-in patch spanning 20$^\circ$ in latitude and longitude.  Silver and blue denote upflow and downflow respectively.  Shown are vantage points from (\textit{a}) above and (\textit{b}) below.}
\label{fig:viz2}
\end{figure*}

\subsection{Subsurface Extrapolation of the Convective Flow}\label{sec:fr}

In order to extrapolate the empirical convective flow field at the surface 
downward, a suitable radial function is necessary which confines the convective 
motions to the upper convection zone, as discussed in Sec.\ \ref{sec:emp}.
To achieve this, choose a function which satisfies the following boundary conditions:
\begin{eqnarray}
f_\ell(r)=0 ~ & \& & ~ f_\ell^\prime(r)=-1, ~~ r=R\label{eq:rcon1} \\
f_\ell(r)=1 ~ & \& & ~~  r=r_w\\
f_\ell(r)=0 ~ & \& & ~ f_\ell^\prime(r)=0,~~ r=r_p \\
f_\ell(r) = 0 & & r \leq r_p \label{eq:rcon4}
\end{eqnarray}
Here $r_p$ is the penetration depth and $r_w = (r_p + R)/2$ is the middle of the convective layer, near the point where $f_\ell(r)$ peaks and $f_\ell^\prime(r)$ changes sign. It can thus regarded approximately as the convective turnover depth (see Fig.\ \ref{fig:radfnc}).   The conditions (\ref{eq:rcon1})--(\ref{eq:rcon4}) ensure that there is no convective mass flux through the surface or through the penetration radius $r_p$.  Furthermore, the normalization ensures that the amplitudes of $W$ and $\tilde{g}_{\ell m}$ are on the order of $\rho(R) U$ where $U$ is a characteristic convective velocity amplitude at the surface.  

These conditions can be satisfied with a fourth-order polynomial:
\begin{equation*}
f(x)=a + bx + cx^2 + dx^3 +ex^4
\end{equation*}
where $x=(r-r_w)/(R-r_w)$, $a=1$, $b=0.25$, $c=-1.75$, $d=-0.25$, and $e=0.75$.  The function $f_\ell(x)$ and its first derivative $f^\prime_\ell(x)$ are shown in Fig.\ \ref{fig:radfnc}.  Note that $x=-1$, $x=0$ and $x=1$ correspond to $r_p$, $r_w$ and $R$ respectively.

\begin{table*}[]
\centering
\caption{Simulation Summary\label{cases}}
\begin{tabular}{c|c|c|c|c|c|c|c|c|c|c|c}
\hline
\hline
Cases\tablenotemark{a} & $\alpha_{spot}$ & $\eta_{top}$ & Cycle & Migration & \multicolumn{2}{|c|}{$B_{pol}$ (G)} & \multicolumn{2}{|c|}{$B_{tor}$ (kG)} & \multicolumn{2}{|c|}{$B_{nax}$ (kG)}& $\frac{B_{tor}}{B_{pol}}$\\
\cline{6-11}
 & & (cm$^2$ s$^{-1}$) & Period (years) & Speed\tablenotemark{b} (m s$^{-1}$) & Mean & $\sigma$&Mean&$\sigma$&Mean&$\sigma$&\\
\hline
 A1 & 15.0 & $3.0\times 10^{12}$ & 13.3 & 11.6 & 596.6 & 118.3 & 33.54 & 1.91 & 1.76 & 0.51 & 56.45 \\ 
 A2 & 25.0 & $1.0\times 10^{13}$ & 13.8 & 8.8  & 67.0 & 13.6  & 4.04 & 0.61 & 0.21 & 0.07 & 60.31 \\ 
 A3 & 25.0 & $3.5\times 10^{13}$ & sub-critical & -- & -- & -- & -- & --  & --   &  --  & --   \\ 
 A4 & 15.0 & $8.0\times 10^{11}$ & 14.0 & 16.6 & 1523.2 & 292.8 & 71.18 & 4.47 & 4.64 & 0.93 & 46.73 \\ 
 A5 & 15.0 & $3.0\times 10^{11}$ & 14.7 & 18.4 & 2075.5 & 435.2 & 84.72 & 9.88 & 6.97 & 1.10 & 40.94 \\ 
 A6 & 15.0 & $1.0\times 10^{11}$ & 15.1 & 19.4 & 2431.4 & 430.1 & 89.24 & 9.58 & 9.37 & 1.15 & 36.68 \\ 
 A7 & 15.0 & $5.0\times 10^{10}$ & 15.1 & 20.1 & 2546.7 & 455.2 & 85.73 & 8.37 & 11.13 & 1.60 & 33.64 \\ 
 A8 & 15.0 & $1.0\times 10^{10}$ & 15.9 & 20.5 & 2531.7 & 388.2 & 89.74 & 12.33 & 12.08 & 1.35 & 35.56 \\ 
\hline
 C1 & 15.0 & $5.0\times 10^{11}$ & 17.4 & 19.1 & 1673.43 & 252.9 & 101.71 & 4.16 & 2.87 & 0.46 & 60.78\\ 
 C2 & 1.0  & $5.0\times 10^{11}$ & sub-critical & -- &--&--&--&--&--&--&--\\ 
 C3 & 15.0 & $5.0\times 10^{10}$ & 17.6 & 18.5 &1800.73 & 293.1 & 112.6 & 19.1 & 3.46 & 0.59 & 62.53 \\ 
\hline
\hline
\end{tabular}
\tablenotetext{1}{Cases A1-A8 include only axisymmetric flows (DR and MC).  Cases C1-C3 also incorporate explicit convection.}
\tablenotetext{2}{See Section \ref{sec:migration}.}
\end{table*}

We have yet to specify the penetration depth $r_p$.  In doing so, it is reasonable to assume that larger-scale motions will penetrate more deeply than smaller-scale motions.  The horizontal length scale $L_h$ of a spectral mode at the surface with total wavenumber $\ell$ is approximately given by
\begin{equation}
\left(\frac{2\pi}{L_h}\right)^2 \sim \frac{\ell (\ell+1)}{R^2} ~~~.
\end{equation}
If we assume that the vertical length scale $L_v$ of a convective mode is comparable to its horizontal length scale, and if we assume that $\ell >> 1$ as it is for most of the convective power (Fig.\ \ref{fig:conhath}), then this gives 
\begin{equation}
L_v \approx L_h \approx \frac{2\pi}{\ell} ~ R  ~~~.
\end{equation}
Thus, we set $r_p = R - L_v = R (1 - 2\pi/\ell)$.  However, we set a minimum value of $r_p = 0.9$ so even the largest motions do not penetrate below this. 

The resulting convective flow fields are illustrated in Figures \ref{fig:viz1} and \ref{fig:viz2}.  The increase of the horizontal length scale of the convection with depth is clear by comparing Fig.\ \ref{fig:viz1}\textit{a} and Fig.\ \ref{fig:viz1}\textit{b} and by comparing the two frames in Fig.\ \ref{fig:viz2}. Fig.\ \ref{fig:viz1}\textit{c} highlights the overturning nature of the motions.  Note that at this resolution, the asymmetry between upflows and downflows is not apparent.  In all calculations reported here we have used radial grid $N_r=340$, latitudinal grid $N_\theta = 512$ and longitudinal grid $N_\phi = 1024$.  

Once this 3D convective flow, ${\bf v}_c$, is defined, we add it to the axisymmetric flows ${\bf v}_a$ such that ${\bf v} = {\bf v}_a + {\bf v}_c$.  Here ${\bf v}_a$ includes the differential rotation and the meridional flow described in Section \ref{sec:STABLE}.  As mentioned in Section \ref{sec:intro}, we restrict our attention in this paper to the case in which ${\bf v}_c(r,\theta,\phi)$ is independent of time.  In future work we will implement an evolving flow field as in the AFT model of \citet{upton14a,upton14b}.

\section{Axisymmetric Flows}\label{sec:nocon}
\begin{figure*}[!t]
\centering
\includegraphics[width=0.95\textwidth, angle=0]{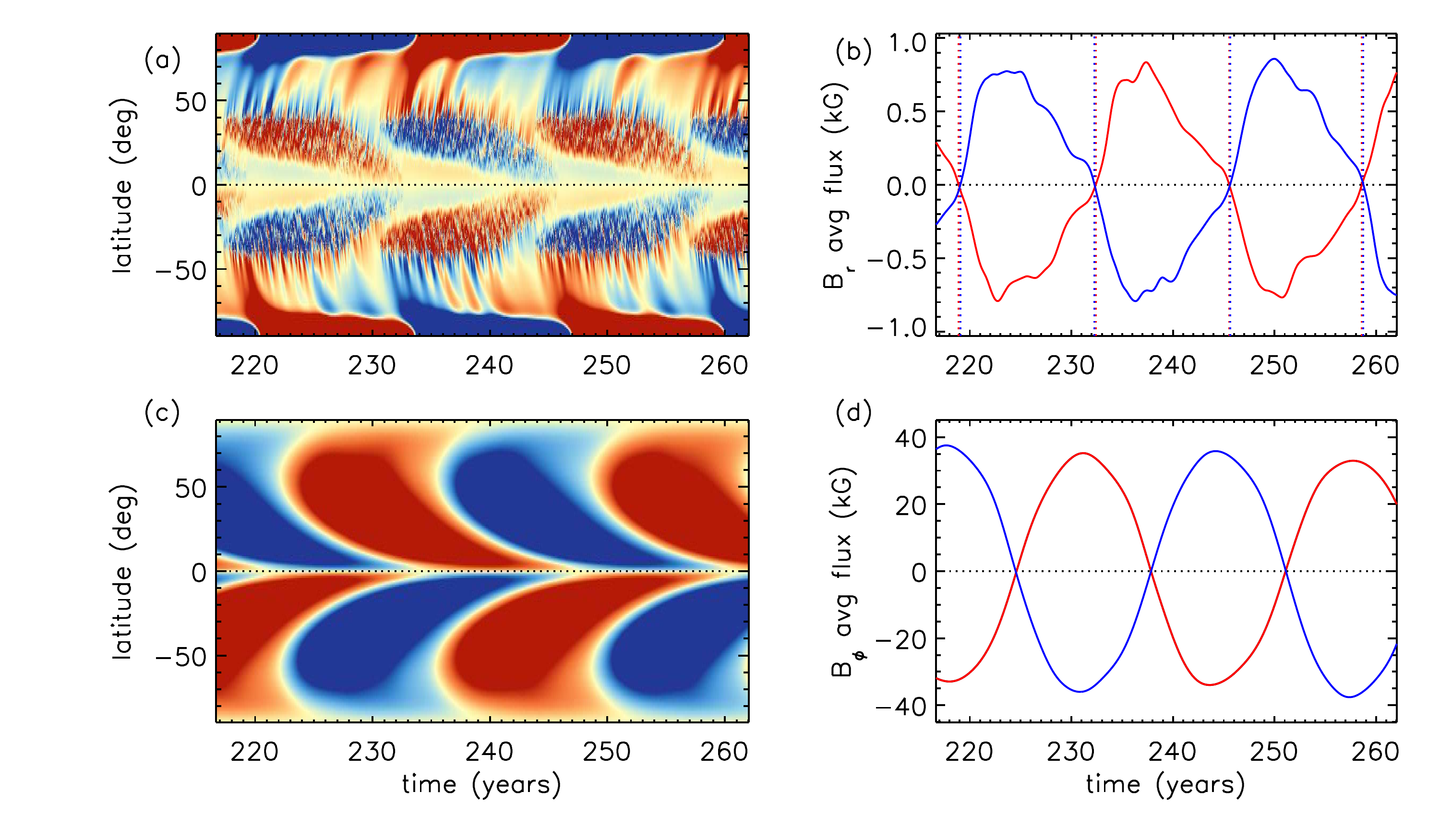}
\caption{(\textit{a}) Time-latitude plot of the mean radial magnetic field on the solar surface, $\left<B_r\right>$, for Case A1. The color scale is set from $-200$G (blue) to +200 G (red). (\textit{b}) Mean polar field calculated by averaging the radial field in (\textit{a}) over latitudes poleward of $\pm 88^\circ$.  Blue and red curves correspond to the northern and southern hemispheres respectively and dotted lines of each color indicate polar field reversals.  (\textit{c}) Time-latitude plot of the mean toroidal field $\left<B_\phi\right>$ at the bottom of the convection zone $r = 0.71 R$.  The color scale saturates at $\pm$ 50 kG, with red and blue denoting eastward and westward field respectively.  (\textit{d}) Mean toroidal flux at low latitudes near the base of the CZ, obtained by averaging the plots in (\textit{c}) over the northern (blue) and southern (red) hemispheres.}
\label{fig:bfly1}
\end{figure*}

In this section, we present some results for the case in which ${\bf v}$ is axisymmetric, consisting only of differential rotation and meridional circulation. These will be compared to the results presented in Section \ref{sec:con} that include 3D convective motions.   All of the simulations presented in this paper are summarized in Table \ref{cases}.  $B_{pol}$ and $B_{tor}$ refer to the amplitude of the mean poloidal and toroidal fields and $B_{nax}$ refers to the amplitude of the non-axisymmetric field components.  In each case we list the mean and standard deviation $\sigma$ of the time series.  The cycle period is calculated by reversals of the mean toroidal field in the lower CZ (e.g.\ Figs.\ \ref{fig:bfly1}\textit{d} and \ref{fig:bfly_con}\textit{d}).  The primary motivation for the range of diffusive simulations A1-A8 is to assess what value of $\eta_{top}$ most closely corresponds to explicit convective transport, as discussed in Sections \ref{sec:con} and \ref{sec:discussion}.  

The initial conditions for all cases is a weak dipole field.  This gets stretched out by the differential rotation and when the toroidal field strength exceeds the threshold value of 1 kG, BMRs begin to appear.  Soon the field strengths become strong enough to saturation the dynamo by means of eq.\ (\ref{flux}).  Case A1 has been run for over 260 years, which corresponds to about 10-15 cycles after the dynamo saturates and settles down to a steady cycle.  Cases A2-A8 have each been run for about 120 years, around 6-7 cycles after saturation.  Cases C1 and C3 have also been run for nearly 190 years each (7-8 cycles after saturation, see Sec.~\ref{sec:con}).

As discussed in Sec.\ \ref{sec:intro}, SFT models rely on flux injection through artificial or observed BMR databases and solve only the radial component of the induction equation. Meanwhile, previous 2D FTD models were not able to reproduce the evolution of the non-axisymmetric components of the surface field.  STABLE unifies these two models by spontaneously producing BMRs based on the dynamo-generated toroidal field and by capturing their subsequent evolution after emergence.  

One of the main aims of STABLE is to reproduce the observed butterfly diagram (radial field) of the solar photosphere.  Though producing solar-like butterfly diagrams is a valuable test of any solar dynamo model, most rely on the mean toroidal field near the base of the convection zone as a proxy for the surface field.  Since STABLE produces explicit BMRs, there is no need for such a proxy.  Both quantities are shown in Fig.\ \ref{fig:bfly1}.  Notable solar-like features include equatorward propagation of active bands at low latitudes and poleward migration of trailing BMR flux at high latitudes, which eventually reverses the polar fields. The equatorward propagation of active bands at the surface (Fig.\ \ref{fig:bfly1}\textit{a}) is a consequence of the subsurface toroidal field propagation (Fig.\ \ref{fig:bfly1}\textit{c}). Note that polar field reversals occur a few years after the peak toroidal field, comparable to solar observations (Fig.\ \ref{fig:bfly1} \textit{c},\textit{d}). 

\begin{figure*}[]
\centering
\includegraphics[width=0.95\textwidth, angle=0]{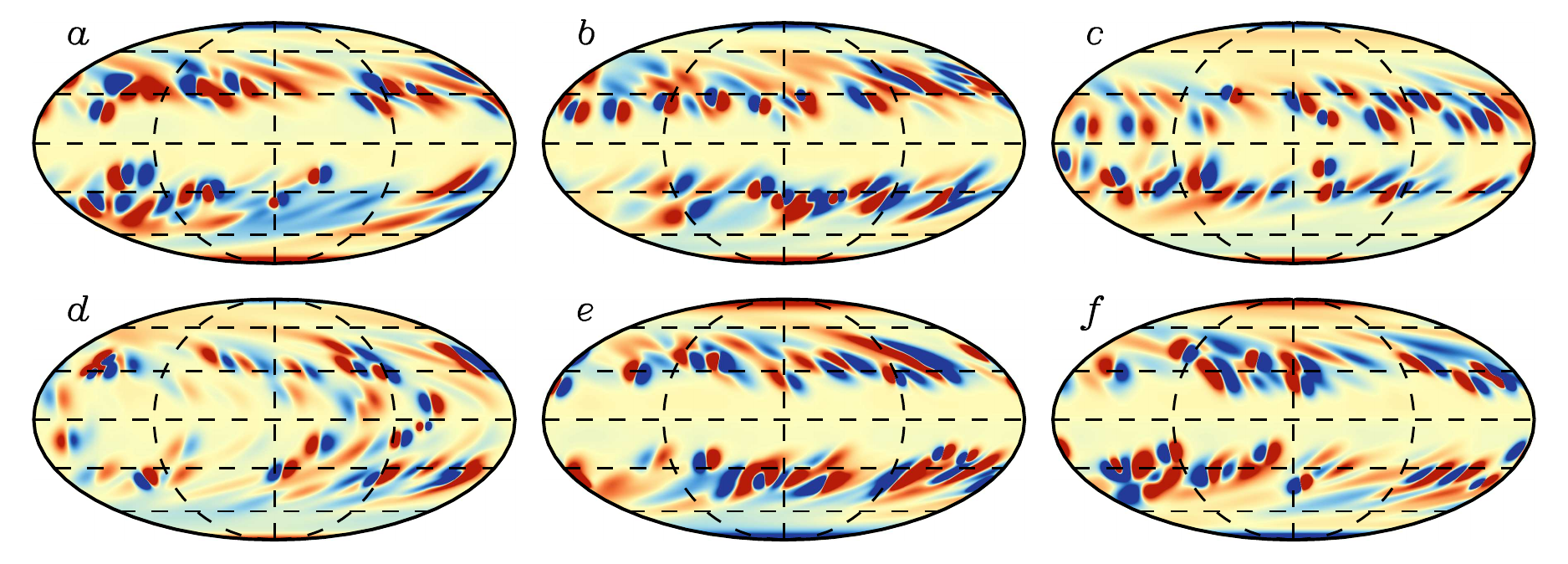}
\caption{Molleweide projection of $B_r$ at $r=R$ in Case A1 at six different times that span a full magnetic cycle: (\textit{a}) 223.9 yr, (\textit{b}) 226.0 yr, (\textit{c}) 229.0 yr, (\textit{d}) 232.0 yr, (\textit{e}) 234.0 yr and (\textit{f}) 237.0 yr. Red and blue denote outward and inward field respectively and the saturation level on the color table is $\pm $ 1 kG.  Dashed lines denote latitudes of 0$^\circ$, $\pm 30^\circ$, and $\pm 60^\circ$.}
\label{fig:ss_A1}
\end{figure*}

\begin{figure*}[]
\centering
\includegraphics[width=0.95\textwidth, angle=0]{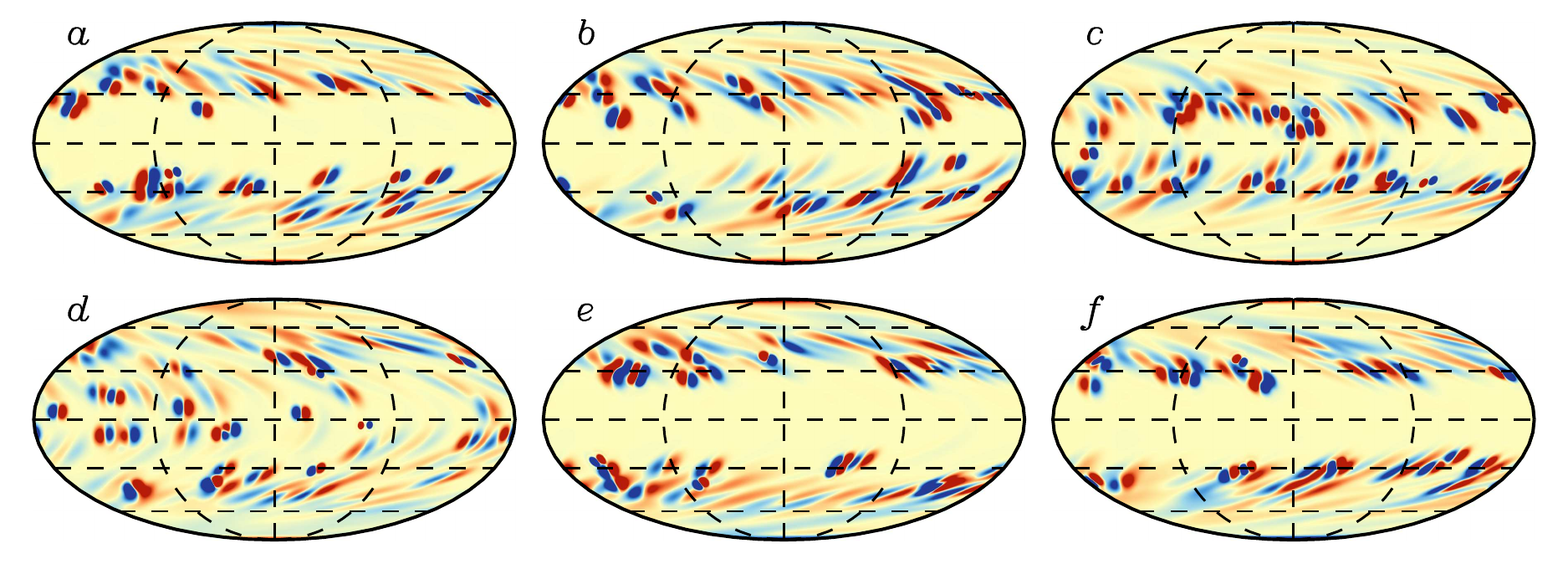}
\caption{Similar to Fig.\ \ref{fig:ss_A1} but for Case A4, with a color saturation level of $\pm$ 3 kG.  The times again span a full magnetic cycle: (\textit{a}) 124.0yr, (\textit{b}) 127.0 yr, (\textit{c}) 130.0 yr, (\textit{d}) 133.0 yr, (\textit{e}) 136.0 yr and (\textit{f}) 139.0 yr.}
\label{fig:ss_A4}
\end{figure*}

\begin{figure*}[]
\centering
\includegraphics[width=1.0\textwidth, angle=0]{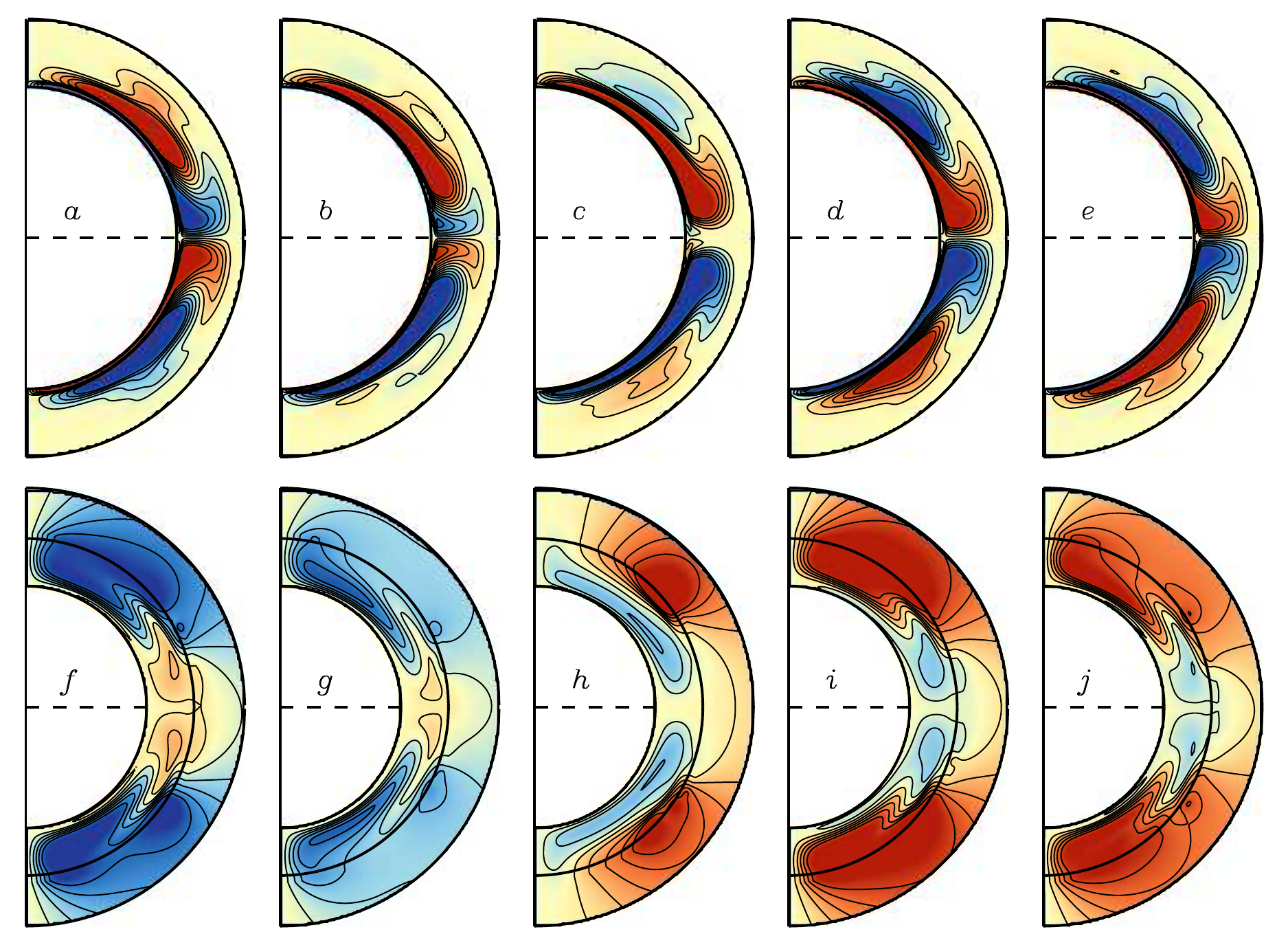}
\caption{Evolution of mean (\textit{a}--\textit{e}) toroidal and (\textit{f}--\textit{j}) poloidal fields in Case A1, spanning one magnetic cycle.  Snapshots are shown for the same magnetic cycle as in Fig~\ref{fig:ss_A1}: $t =$ (\textit{a},\textit{f}) 226.0 yr, (\textit{b},\textit{g}) 229.2 yr, (\textit{c},\textit{h}) 233.0 yr, (\textit{d},\textit{i}) 237.1 yr and (\textit{e},\textit{j}) 239.9 yr.  Frames (\textit{a}--\textit{e} show $\left<B_\phi\right>$ with red and blue indicating eastward and westward field respectively.  Frames (\textit{f}--\textit{j}) show the poloidal magnetic potential with a potential field extrapolation above $r=R$ to $r=1.25 R$.  Colors indicate (red) clockwise and (blue) counter-clockwise field orientations.  Average toroidal and poloidal field strengths are about 33 kG and 596 G respectively, as listed in Table \ref{cases}, though peak toroidal fields can exceed 100 kG (the color table for $\left<B_\phi\right>$ is clipped at $\pm 50$ kG).}
\label{fig:bmean_sand1}
\end{figure*}

Figure \ref{fig:ss_A1} highlights the surface flux transport in Case A1.  Compare this with Fig.\ \ref{fig:ss_A4}, which shows the same thing but for a case with a lower surface diffusion (Case A4).  The structures in the former case tend to be wider and spread out, commensurate with the higher diffusion. This is reflected in lower average field strengths for the non-axisymmetric field components: 1.76 kG in Case A1 versus 4.64 kG in Case A4 (Table \ref{cases}). Higher diffusion always make a dynamo less efficient. The mean fields in Case A5-A8 are about 3-4 times stronger than in Case A1 even though the quenching field strength $B_q$ is the same in both Cases (see eq.\ \ref{flux}).  The evolution of the mean fields for Case A1 is shown in Fig.\ \ref{fig:bmean_sand1}, which highlights the general FTD aspects of these simulations. The basic operation of the dynamo is similar to that described in \citet{MD14} and \citet{MT16}. Differences between those previous results and the results shown here are mainly due to modified meridional circulation profile described in \citet{HCM17} and the varying values of $\eta_{top}$.

The cycle period increases somewhat as $\eta_{top}$ is decreased (Table \ref{cases}).  We attribute this to a ``short-circuiting'' of the flux-transport dynamo by diffusive mixing.  Even though all cases have the same value of $\eta$ in the mid and lower CZ, a greater value of $\eta_{top}$ promotes a more efficient downward transport of poloidal flux from the surface to the base of the CZ.  The difference is significant, but not substantial.  A decrease in $\eta_{top}$ by a factor of 300 (from A1 to A8) lengthens the cycle period by about 20\%.  

Note that this result is in contrast to other FTD dynamo parameter studies that report a slight increase in the cycle period as the diffusion is increased \citep{DC99,Yeates08}.  However, these studies mainly focus on increasing the diffusivity in the bulk of the convection zone, not in the surface layers.  For example, in terms of our notation, \citet{Yeates08} vary $\eta_{mid}$ while keeping $\eta_{top}$ fixed, whereas we do the opposite.  This accounts for the difference in our results, as we now explain.

There are at least three ways in which an increase in the turbulent diffusion can influence the cycle period in a Flux-Transport dynamo model:  (A) The ``short-circuit'' effect noted above, by which diffusion enhances the efficiency of poloidal flux transport across the CZ, (B) a decrease in the efficiency of equatorward transport of toroidal flux at the base of the CZ, and (C) a decrease in the efficiency of toroidal flux generation at the base of the CZ.  Effect (A) tends to decrease the cycle period, as described above.  Meanwhile, effects (B) and (C) tend to increase the cycle period.  

Effect (B) arises because of a decrease in the magnetic Reynolds number associated with the MC near the base of the CZ, Rm $ = U_{mc} L_{mc}/\eta_{mid}$, where $U_{mc}$ and $L_{mc}$ are the relevant velocity and length scales.  Lower Rm inhibits transport, as the magnetic field slips through the plasma.  This slows down the equatorward advection of both toroidal and poloidal flux.  It also tends to smooth out the structure of the field, which decreases the field amplitudes and inhibits toroidal field generation by the $\Omega$-effect.  This is effect (C) referred to in the previous paragraph.  In a purely kinematic model, the toroidal field generation time scale would be independent of the field strength.  However, many parameter studies such as \citet{DC99} and \citet{Yeates08} include a quenching field strength for the toroidal field which limits the amplitude of the dynamo.  At high values of $\eta_{mid}$, it takes the dynamo longer to reach this quenching field strength.  

The simulations by \citet{DC99} and \citet{Yeates08} include all three effects, but the latter two dominate.  For example, a comparison of Figs.\ 8\textit{d} and 8\textit{i} in \citet{Yeates08} clearly shows effect (A).  By contrast, our simulations only include effect (A) because we keep the value of $\eta_{mid}$ (and thus the value of Rm in the lower CZ) fixed in all of our simulations.  We are aware of one previous study that investigated the influence of $\eta_{top}$ on the operation of an FTD dynamo with fixed $\eta_{mid}$, namely \citep{Hotta10}.  They find, as do we, that the cycle period decreases with increasing $\eta_{top}$ (see their Fig.\ 5).

\section{Convective Flows}\label{sec:con}

In this section we describe dynamo simulations that include 3D convective flow fields as described in Section \ref{sec:cflow}.  We remind the reader that these are not full MHD simulations.  Rather, the convective velocity field is prescribed in a kinematic sense based on observations of the photospheric power spectrum.  We also remind the reader that the imposed convection does not occupy the entire convection zone.  Rather, it is intended to mimic the relatively vigorous, small-scale convective motions in the solar surface layers ($r > 0.9R$) that are well understood both observationally and theoretically. This imposed convective flow is in lieu of the turbulent diffusion that is used to represent surface convection in most SFT and FTD models \citep{Jiang_review15}.  Thus, we correspondingly reduce the turbulent diffusion coefficient near the surface, $\eta_{top}$, as shown in Fig.\ \ref{fig:eta}.  However, $\eta_t$ is still large enough to suppress small-scale dynamo action (\ref{sec:ssd}) and to promote dipolar parity (\ref{sec:transport}.  So, it is of the same order as A5-A8 (see Table \ref{cases}).  The AFT model employs a similar value for the background $\eta_t$.  In the deeper CZ we continue to represent convective transport as a turbulent diffusion.  For further details on the implementation, see Section \ref{sec:cflow}.

\subsection{Amplification of fields by convective flows}\label{sec:ssd}
Babcock-Leighton (BL) dynamo models rely on the dispersal and migration of magnetic field by near-surface convection but they typically neglect the fact that these same convective motions can operate as a small-scale dynamo \citep[e.g.][]{Cattaneo99}.   In our model, when we include 3D convective motions as described in Sec.\ \ref{sec:cflow}, we find that this flow does indeed operate as a small-scale dynamo.

A characteristic length scale of the convection can be estimated as $L \sim 2 \pi R / \ell_0 \sim$ 34 Mm, where $\ell_0 \sim 130$ is the spherical harmonic degree at which the spectrum peaks (Fig.\ \ref{fig:conhath}\textit{c}).  For a velocity scale, we can use the rms value of $U \sim 250$ m s$^{-1}$.  Together with $\eta_{top} = 5 \times 10^{10}$ cm$^2$ s$^{-1}$, this implies a magnetic Reynolds number of $R_m = UL/\eta_{top} \sim 1700$.  So, it is not surprising that this flow is supercritical to small-scale dynamo action.  We find that it is even supercritical for $\eta_{top} = 5 \times 10^{11}$ ($R_m \sim 170$).  Though this behavior is reasonable, it is not conducive to establishing magnetic cycles.

We find that small scale dynamo action disrupts the operation of the BL dynamo.  BMRs are quickly overwhelmed by random small-scale fields that grow exponentially. This inhibits poloidal field generation by the BL mechanism and thus suppresses the magnetic cycles.  This problem is not found in the AFT model \citep{upton14a,upton14b}, which uses similar surface flows (though time evolving), because AFT is 2D and therefore cannot exhibit sustained dynamo action.

We have considered several approaches to suppressing this small-scale dynamo action in order to get a functioning BL dynamo.  First, we enhanced the dissipation on small (convective) scales by imposing a horizontal hyperdiffusion $\nabla_h^4$ or $\nabla_h^8$.  But we were unable to find appropriate hyperdiffusion coefficients to yield a cyclic dynamo solution. Then, we attempted to saturate the small-scale fields by introducing a threshold field strength $B_s$, applied to the non-axisymmetric field components.  However, it was not possible to apply this selectively to suppress dynamo-generated fields but not the remnant fields from emerging BMRs.  We also tried nonlinear quenching of the non-axisymmetric field, with similar results.

We were able to achieve some degree of BL activity by using a drag term of the form
\begin{equation}
\frac{{\pd \bf B}^\prime}{\pd t} = - \frac{{\bf B}^\prime}{\tau} + \ldots
\end{equation}
where ${\bf B}^\prime$ is the non-axisymmetric field and $\tau \sim$ 24 hrs is a suppression time scale that is comparable to the small-scale dynamo growth rate.   This is effectively a scale-independent variation of the hyperviscosity approach.  However, the solutions were not solar-like, exhibiting a large asymmetry about the equator, with BMRs in one hemisphere and not the other.

The only effective way we found to suppress small-scale dynamo action is by making the convective flows operate only on the radial component of the magnetic field.  This allows us to realistically capture the horizontal transport of vertical magnetic flux in the solar surface layers, which is the most important component of the turbulent transport from the perspective of the Babcock-Leighton paradigm.  Other tensorial components of the turbulent diffusivity and magnetic pumping would be interesting to capture but they are of secondary importance and they can await future work.  Thus, for Cases C1, C2, and C3 presented in this paper, we have only applied the convective flows to the radial field $B_r$ only. And, even in this case, we still found it beneficial to employ a relatively strong magnetic diffusion in order to suppress small-scale fields (see Sec.\ \ref{sec:STABLE}).  Though the value of $\eta_t$ (5$\times 10^{10}$ -- 5$\times 10^{11}$ cm s$^2$) is comparable to that used in some of the diffusive cases (see Table \ref{cases}), the value of the hyperdiffusion is substantially larger ($\eta_h = 2\times 10^{10}$ cm$^2$ s$^{-1}$, compared to $2 \times 10^8$ cm$^2$ s$^{-1}$ used for Cases A1-A8).  In this way we were able to achieve a viable BL dynamo model as described in section \ref{sec:transport}. 

We appreciate that this solution is not ideal, but it is required by the kinematic nature of our model.  Capturing both the surface flux transport and the small-scale dynamo action in the solar surface layers will ultimately require a more sophisticated MHD model with Lorentz-force feedbacks.  Since this would require a substantial model development, we defer it to a future publication.

\begin{figure*}[]
\centering
\includegraphics[width=0.95\textwidth, angle=0]{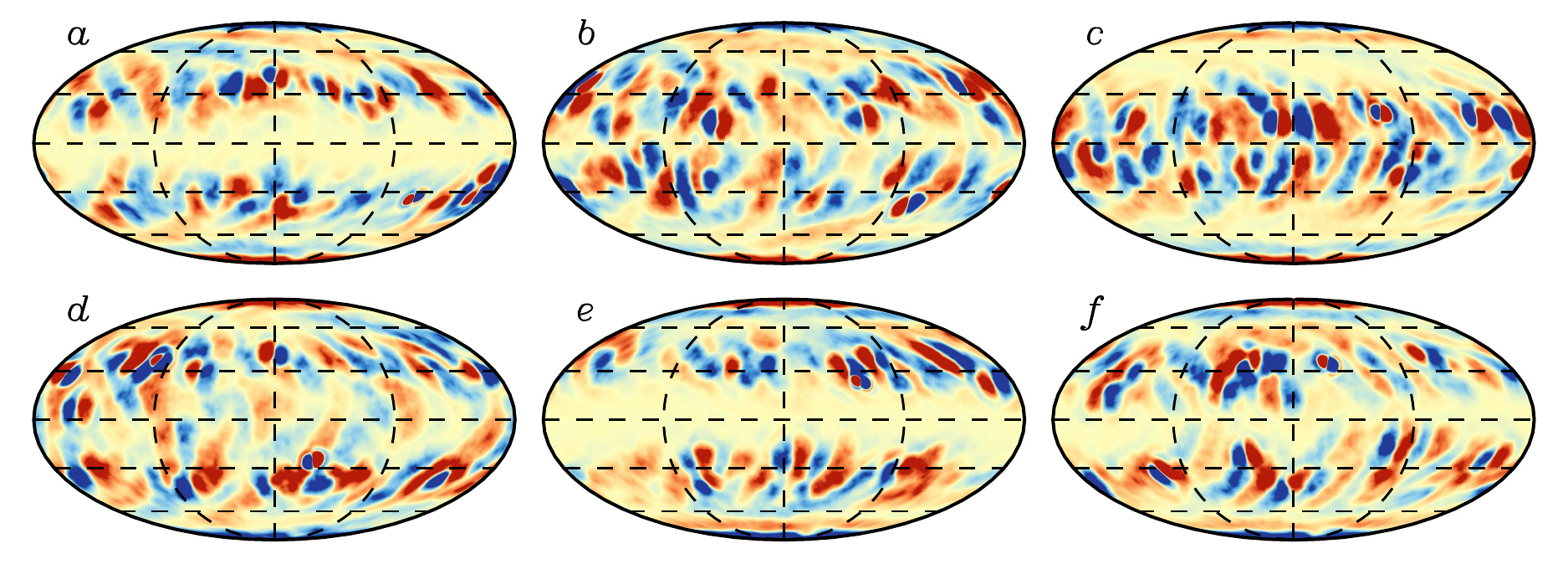}
\caption{Similar to Figures \ref{fig:ss_A1} and \ref{fig:ss_A4} but for the convective Case C1, with a color saturation level of $\pm$ 3 kG.  Snapshots are shown for $t =$ (\textit{a}) 140.0 yr, (\textit{b}) 144.0 yr, (\textit{c}) 148.0 yr, (\textit{d}) 152.0 yr, (\textit{e}) 156.0 yr and (\textit{f}) 160.0 yr.}
\label{fig:ss_con}
\end{figure*}

\begin{figure*}[t!]
\centering
\includegraphics[width=0.95\textwidth, angle=0]{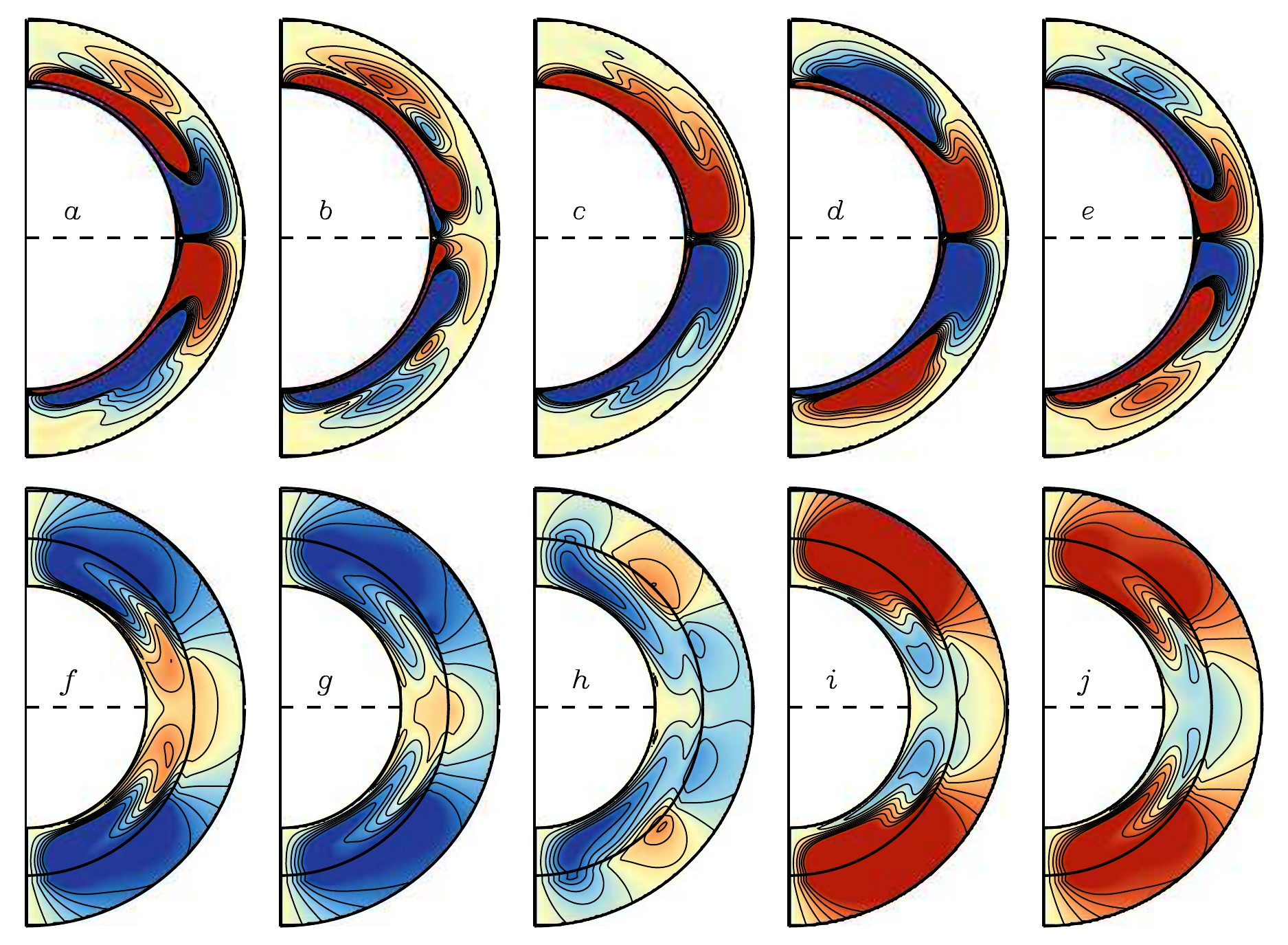}
\caption{As in Fig.\ \ref{fig:bmean_sand1} but for Case C1.  This spans the same magnetic cycle as in Fig.\ \ref{fig:ss_con}, at times of $t = $ (\textit{a},\textit{f}) 140.0 yr, (\textit{b},\textit{g}) 145.0 yr, (\textit{c},\textit{h}) 150.0 yr, (\textit{d},\textit{i}) 155.0 yr and (\textit{e},\textit{j}) 160.0 yr.}
\label{fig:bmean_con}
\end{figure*}

\begin{figure*}[t!]
\centering
\includegraphics[width=0.95\textwidth, angle=0]{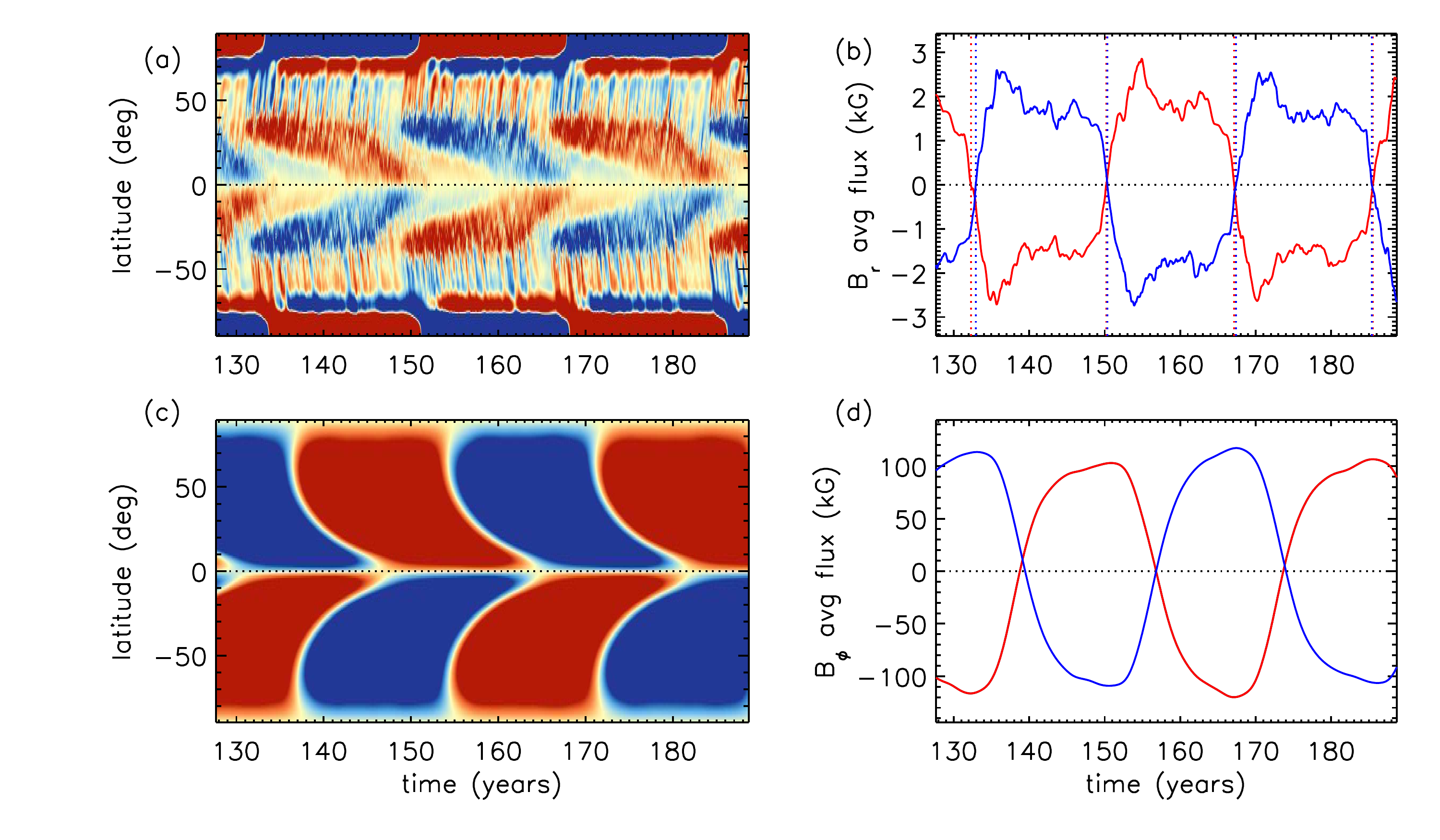}
\caption{As in Fig.\ \ref{fig:bfly1} but for Case C1.  Here the color tables saturate at (\textit{a}) $\pm $ 500 G and (\textit{c}) $\pm$ 100 kG.}
\label{fig:bfly_con}
\end{figure*}

\subsection{Results with convective flows}\label{sec:transport}
In this section, we present results with explicit convective motions as the effective transport mechanism of large scale magnetic field on the surface of the Sun, in addition to meridional circulation and differential rotation.  This convective transport is limited to the horizontal advection of vertical field near the upper convection zone as described in Sec.\ \ref{sec:ssd}, though we still decrease the explicit diffusion $\eta_{top}$ as illustrated in Fig.\ \ref{fig:eta}.  Thus, when we refer to including {\em convective flows} in our models we really mean explicit convective flow fields extrapolated upto $0.9R_\odot$ from the surface, chosen to mimic the observed properties of convection near the solar surface.  Convective transport in the mid convection zone is still modeled with a turbulent diffusivity.

It is worth noting that the introduction of the convective flow field ${\bf v}_c$ increases the computational expense substantially. This is because of the Courant-Friedrichs-Lewy (CFL) constraint on the time step.  At the horizontal resolution used here ($N_\theta = $ 512, $N_\phi = 1024$), the peak convective velocity is on the order of 900 m s$^{-1}$.  This is a factor of 6.7 larger than the peak axisymmetric flow speed of 135 m s$^{-1}$, requiring a commensurate decrease in the time step.  If vertical convective motions are included this requires a further decrease of the time step since the vertical grid spacing is about a factor of 7 smaller than the horizontal grid spacing.  Turbulent diffusion is not subject to a CFL constraint because it is handled by a semi-implicit Crank-Nicolson scheme.

We have performed three simulations with convection, as summarized in Table \ref{cases}.  Case C1 is a fiducial solar-like dynamo solution that we will discuss further below. Case C2 is similar to Case C1 but with $\alpha_{spot} = 1$.  This is found to be sub-critical for sustained dynamo action, as found in many diffusive cases (MT16).  Case C3 is similar to Case C1 but with a lower value of $\eta_{top}$; $5 \times 10^{10}$ cm$^2$ s$^{-1}$ instead of $\eta_{top}$; $5 \times 10^{11}$ cm$^2$ s$^{-1}$.  The rationale for choosing a low value is because $\eta_{top}$ is intended to parameterize the convective transport that we are now capturing explicitly, as emphasized in Sec.\ \ref{sec:cflow}.  However, we find that if we make $\eta_{top}$ too low the dynamo flips into a quadrupolar parity ($\left<B_r\right>$ and $\left<B_\phi\right>$ symmetric about the equator).  We believe that Case C3 is in that quadrupolar regime.  Though it's parity is still largely negative (dipolar) after 190 years of evolution, it is tending toward quadrupolar. Furthermore, a longer simulation with lower resolution (otherwise the same parameters) shifts to quadrupolar after about 300 years.

This is a known feature of BL dynamo models.  Efficient diffusive coupling between the two hemispheres is known to promote dipolar parity ($\left<B_r\right>$ and $\left<B_\phi\right>$ antisymmetric about the equator), as demonstrated by \citet{CNC04} and \citet{Hotta10}.  Evidently, the convective transport alone is not sufficient to establish dipolar parity in our models. In short, the net turbulent transport (advection plus diffusion) in Case C3 appears to be insufficient to yield dipolar parity but in Case C1 it is.  The value of $\eta_{top}$ in Case C1 is high enough to enhance the subsurface diffusive coupling between hemispheres but low enough that the explicit convective flows dominate the surface flux transport (as will be demonstrated below).  Since the Sun exhibits dipolar parity, we will hereafter focus on Case C1.

Fig.\ \ref{fig:ss_con} shows the evolution of the surface flux in Case C1.  Qualitatively it looks more like Case A1 (Figs.\ \ref{fig:ss_A1}) than A4 (Figs.\ \ref{fig:ss_A4}), with more diffuse bipolar structures at low latitudes, though Case C1 has stronger fields (Table \ref{cases}).   The evolution of the mean fields in Case C1 is also similar to Case A1, as shown in Fig.\ \ref{fig:bmean_con}.   This suggests that the efficiency of convective flux transport is comparable to that of the turbulent diffusion in Case A1. We will see below that this first impression is borne out by a more thorough analysis. 

The poloidal and toroidal butterfly diagrams for Case C1 are shown in Fig.\ \ref{fig:bfly_con}.  The most apparent differences in the surface flux transport (Fig.\ \ref{fig:bfly_con}\textit{a}) relative to Case A1 (Fig.\ \ref{fig:bfly1}\textit{a}) are less pronounced bipolar structures in the low latitudes and mix polarity polar caps. In Case C1, active regions are shredded by convective flows which results a less concentrated active regions compared to Case A1. A more subtle difference is that the poleward migration rate of the streams is somewhat more rapid (see Section \ref{sec:migration}).  Though the broader polar regions in Case C1 possess mixed polarities, the polar fields are stronger than the unipolar polar regions of Case A1; compare Figs.\ \ref{fig:bfly1}\textit{b} and \ref{fig:bfly_con}\textit{b}.  Such behavior may be attributed to the tendency for the convective motions to disperse and transport BMR fields without dissipating them (see Sec.\ \ref{sec:energetics}).  In the case of A1, BMRs emerge and the opposite polarities partially cancel each other as they disperse.  By contrast, in Case C1, the fields disperse but cancellation is less efficient as a result of the smaller ohmic diffusion.  Thus, both polarities are transported poleward and concentrated into strong, alternating bands.

The shape of the polar flux plot for Case C1 (\ref{fig:bfly_con}\textit{b}) is similar to Case A1 (Fig.\ \ref{fig:bfly1}\textit{b}) but with somewhat more variation and a sharper decay at the end of each cycle.  This variability reflects the mixed polarity fields that cross into the polar regions before they cancel one another. The slower decay phase for Case C1 implies a longer interval of polar flux generation by poleward migrating streams which persists for almost the entire cycle, as seen in Fig.\ \ref{fig:bfly_con}\textit{a}. By contrast, poleward migrating streams are less prominent late in the cycle for Case A1 (Fig.\ \ref{fig:bfly1}\textit{a}).   As BMRs emerge at progressively lower latitudes, most of the emerging flux in this more diffusive case cancels locally before it can migrate to higher latitudes.  This sustained flux emergence at mid latitudes is not supported by solar observations, which show few mid-latitude BMRs in the declining phase of the cycle. This discrepancy may be due in part to our simple flux emergence algorithm.  Resolving it will likely require a better understanding of the flux emergence process.

The sustained supply of poloidal flux to the poles throughout most of the cycle in Case C1 also has consequences for the toroidal field generation.  As this flux is transported to the base of the CZ by meridional circulation and turbulent diffusion, it promotes sustained toroidal flux generation through the $\Omega$-effect, particularly at mid-latitudes where the latitudinal shear is strongest. This is evident in Fig.\ \ref{fig:bfly_con}\textit{c} which shows strong toroidal fields persist near $\pm 50-60^\circ$ throughout most of the cycle. Compare this to Case A1, (Fig.~\ref{fig:bfly1}\textit(c)), where the mid-latitude toroidal flux diminishes late in the cycle as the bands propagate equatorward.  This excess of mid-latitude flux in Case C1 also accounts for the distortion of the integrated toroidal flux curve (Fig.\ \ref{fig:bfly_con}\textit{d}), which peaks later in the cycle than Case A1 (Fig.\ Fig.\ \ref{fig:bfly1}\textit{d}).  

\begin{figure*}[t!]
\centering
\includegraphics[width=0.95\textwidth, angle=0]{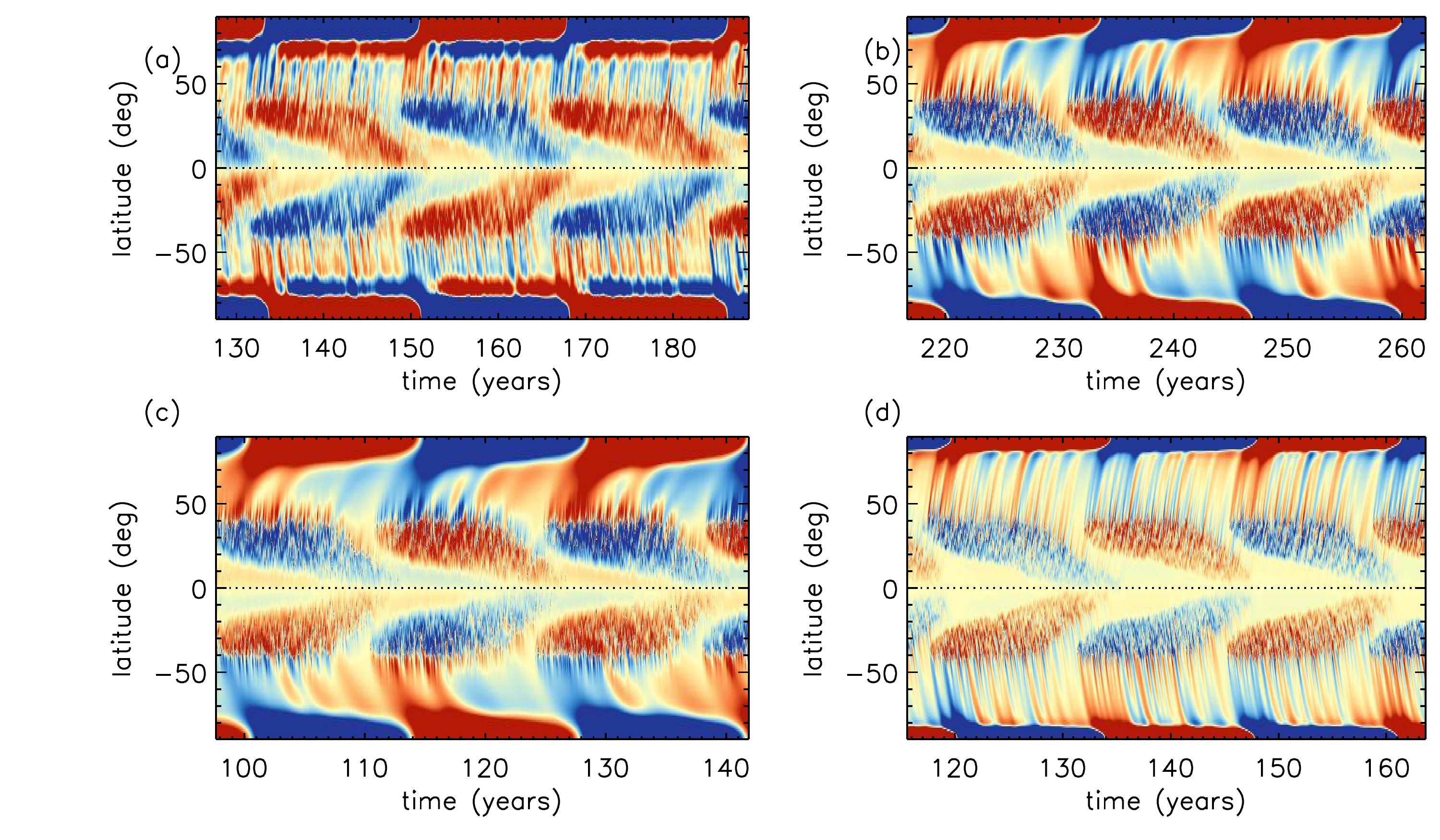}
\caption{Time latitude plot of $\left<B_r\right>$ at $r=R$ for: (a) Case C1 ($B_r$ scale $=\pm 500$G). (b) Case A1 with $\eta_{top} = 3\times 10^{12}$ ($Br$ scale $=\pm 200$ G) (c) Case A2 with $\eta_{top} = 10^{13}$ ($Br$ scale $= \pm 30$ G)  and (d) Case A4 with $\eta_{top} = 8\times 10^{11}$ ($B_r$ scale $=\pm 800$ G).} 
\label{fig:all_bfly}
\end{figure*} 

Figure \ref{fig:all_bfly} compares the surface butterfly diagram in Case C1 to several diffusive cases with relatively high and low values of $\eta_{top}$, ranging from $1 \times 10^{13}$ cm$^2$ s$^{-1}$ to $8 \times 10^{11}$ cm$^2$ s$^{-1}$.  Qualitatively, it bears the greatest resemblance to Case A1. This conclusion is based on the width of the poleward-migrating streams (though note the different ranges for the time axes), the structure and width of the polar flux concentrations, the relative strength of the polar and low-latitude fields, and the location of the active latitudes.  In Section \ref{sec:migration} we take a more quantitative approach to comparing the poleward migration speeds.

It is worth noting that the polar field strength of $\sim$ 2.0 kG in Case C1 is significantly higher than typical measurements of the polar fields in the solar photosphere, which are on the order of 1-3 G \citep[e.g.]{munoz12}. In our model, we calculate the polar field by averaging the radial magnetic field over the region $88^\circ$ to pole.  However, observational measurements usually quote the line-of-sight field strength averaged over a large region, often down to latitudes of 70$^\circ$ or even 55$^\circ$.  Projection and averaging effects therefore diminish the observed field strengths.  After correcting for these effects, the observed value of the polar field strength is on the order of 10 G, which is consistent with the backwards extrapolation of coronal and heliospheric measurements and models \citep{gibso99}.  Also, it must be remembered that the BMR fluxes that we use here are artificially large.  As discussed in Section \ref{sec:STABLE}, we enhance the photospheric flux budget by setting $\alpha_{\rm spot} > 1$ in order to ensure that the dynamo solution is super-critical.  For Case C1 $\alpha_{\rm spot} = 15$.  We have confirmed that setting $\alpha_{\rm spot} = 1$ leads to decaying solutions (Case C2 in Table \ref{cases}).  Still, the high polar field strengths are a concern and are a known issue with BL/FT models.   A possible remedy might be to take into account the connectivity of emerging BMRs with their deep-seated roots.  \citet{YM13} show that this can shift the region of poloidal field generation more to the interior, producing weaker poloidal fields at the surface and particularly at the poles.

In summary, the convective Case C1 behaves in many ways like the diffusive Case A1, which has $\eta_{top}$ $\sim 3 \times 10^{12}$ cm$^2$ s$^{-1}$. However, the mean fields in Case C1 are much stronger.  In the next section we investigate why this is and we take a closer look at the similarities and differences between Case C1 and the cases in which convective transport is approximated by turbulent diffusion.

\section{Discussion: Does Convection Operate as a Turbulent Diffusion?}\label{sec:discussion}

The main objective of our paper is to improve the fidelity of the surface flux transport in our 3D Babcock-Leighton dynamo model by replacing turbulent diffusion with a more realistic depiction of photospheric convection.  However, as mentioned in Section \ref{sec:con}, this introduces challenges, including small-scale dynamo action and reduced computational efficiency.  For this reason, and also for understanding the nature of convective transport, it is important to ask how much we gain from the explicit convective flows.  Is convective transport accurately parameterized by a turbulent diffusion or is is fundamentally non-diffusive?  If the former, what value of $\eta_{top}$ is optimal?  These are the questions we address in this section.

Before proceeding, we emphasize that we do not intend to investigate the fundamental physics of turbulent magnetic diffusion.  This is best done in isolation, with idealized numerical experiments free from complicating factors such as mean flows.  Rather, we {\em are} interested in how explicit convective flux transport behaves within the context of a Babcock-Leighton dynamo model.  All previous Babcock-Leighton models have used turbulent diffusion and, occasionally, turbulent pumping to represent convective transport.  We would like to know whether or not this is a good approximation.

In Section \ref{sec:con} we argued that, at least qualitatively, the convective Case C1 resembles Case A1, which has $\eta_{top} = 3 \times 10^{12}$ cm$^2$ s$^{-1}$.  This is comparable to estimates of the turbulent diffusion from photospheric observations of magnetic flux elements, which suggest $\eta_{top} \sim$ 2--6 $\times 10^{12}$ cm$^2$ s$^{-1}$ \citep{mosher77,topka82,schri96,abram11}.  In mean-field theory, the value of $\eta_t$ is linked to the kinetic energy of the turbulence.  In particular, if $U$ and $L$ are characteristic velocity and length scales of ${\bf v}^\prime$, then $\eta_t \sim U L / 3$ \citep[e.g.][]{ossen03}.  Here we have $U \sim 250$ m$^{-1}$ and $L \sim$ 34 Mm (Sec.\ \ref{sec:ssd}), which suggests a value of $\eta_{MFT} \sim 10^{13}$ cm$^2$ s$^{-1}$ - somewhat higher than the observational estimate.

In the remainder of this section, we take a closer look at these estimates.

\subsection{The turbulent electromotive force (emf)}\label{sec:emf}

In kinematic mean-field dynamo theory, turbulent diffusion is only one component to the turbulent electromotive force (emf), $\emf$.  Though more general averaging procedures are sometimes used, here we define the turbulent emf as
\begin{equation}\label{eq:emf}
\emf = \left<{\bf v}^\prime \cross {\bf B}^\prime\right> 
\end{equation}
where the angular brackets indicate averages over longitude and primes indicate non-axisymmetric components, e.g.\ ${\bf v}^\prime = {\bf v} - \left<{\bf v}\right>$.  

The ansatz of turbulent diffusion as expressed in eq.\ (\ref{induction1}) assumes that
\begin{equation}\label{eq:demf}
\demf \approx -\eta_t \curl \left<{\bf B}\right> ~~~~.
\end{equation}
This is the hypothesis we wish to test. Please note that this is a very crude approximation to parametrize the turbulent emf in terms of mean magnetic field only. There will be others terms corresponding to the rotation and density stratification \citep{kitchatinov94} but we only want to do order of magnitude comparison. Furthermore, since the imposed convective velocity field is non-helical and only operates on the vertical component of ${\bf B}^\prime$, the $\alpha$-effect, the magnetic pumping, and the off-diagonal components of the turbulent diffusivity tensor should all vanish.  So, we can compute $\emf$ explicitly in Case C1 from the non-axisymmetric components of ${\bf v}$ and ${\bf B}$ and then meaningfully ask whether or not it is essentially diffusive in nature as expressed in eq.\ (\ref{eq:demf}).  

If we focus on the longitudinally-averaged radial field $\left<B_r\right>$ at the surface ($r=R$), then the relevant component of the emf in eq.\ (\ref{eq:emf}) is ${\cal E}_\phi = -\left<V_\theta^\prime B_r^\prime\right>$.  When comparing this to the diffusive parameterization in eq.\ \ref{eq:demf}, we wish to focus on the component of $\demf$ that captures the horizontal diffusion of vertical field at the surface, which is
\begin{equation}\label{eq:semf}
{\cal D}_\phi \approx \frac{\eta_t}{R}\left(\frac{\partial \left<B_r\right>}{\partial \theta}\right)_{r=R} ~~~.
\end{equation}     

We find that a value of $\eta_t \sim 2 \times 10^{13}$ cm$^2$ s$^{-1}$ gives similar amplitudes for ${\cal E}_\phi$ and ${\cal D}_\phi$. This is quite similar to the mean-field estimate of $10^{13}$ cm$^2$ s$^{-1}$ given at the beginning of Section \ref{sec:discussion}. However, we found little correlation between ${\cal E}_\phi$ and ${\cal D}_\phi$. Averaging each over a one-month interval in the early and late phases of a typical cycle yields correlation coefficients of about 0.25 and 0.23 respectively.

To appreciate why this correlation coefficient is so low, consider that ${\cal E}_\phi$ is largest where $B_r^\prime$ is largest, namely regions of unipolar field within BMRs and in the polar caps.  Here the convective flow tends to advect vertical magnetic field into regions of horizontal convergence--the equivalent of downflow lanes, though $v_r^\prime$ is effectively zero in Case C1 (see Sec.\ \ref{sec:ssd}).  Thus, $B_r^\prime$ is squeezed before it is dispersed.  This squeezing action is anti-diffusive and opposes the background diffusion, which is weak but non-negligible on the scale of the convergence lanes ($\eta_{top} = 5 \times 10^{11}$ cm$^2$ s$^{-1}$).  

This mimics the evolution of vertical magnetic flux in the solar photosphere, which is anti-diffusive at early times, as the flux is advected toward supergranular boundaries to form the magnetic network, and diffusive at spatial and temporal scales much larger than those of the convective motions \citep{cadavid99}.  This long-term diffusive behavior is difficult to capture by means of the pointwise comparison of ${\cal E}_\phi$ and ${\cal D}_\phi$.  More sophisticated techniques are needed to quantify the effective diffusion coefficient (because of convective flows only), such as correlation tracking of magnetic flux elements.  In the next Section (Sec.\ \ref{sec:migration}) we consider an alternative approach, which is to quantify the speed at which residual BMR flux is transported to the poles.

\subsection{Poleward Migration}\label{sec:migration}

One way to quantify the efficiency of surface flux transport is by estimating the rate at which trailing flux migrates from mid-latitudes to the poles.  In particular, we see that the butterfly diagrams in Figures \ref{fig:bfly1}\textit{a}, \ref{fig:bfly_con}\textit{a} and \ref{fig:all_bfly} are dominated by successive streams of mixed polarities that reflect the propagation of radial field from latitudes below about $\pm 40^\circ$ to latitudes above $\pm 70^\circ$.  How do these propagation/migration rates vary among the different cases?  In this section we address that question quantitatively by means of a correlation function.

We proceed by first constructing a 2D map of the mean, radial, surface magnetic field $\left<B_r\right>(r=R,\theta, t)$ for each case as in Fig.\ \ref{fig:all_bfly} but now limited to the latitude range between 40$^\circ$ and 70$^\circ$.  For shorthand, we will refer to this 2D map as $B(\theta_k,t)$ where $k$ is the discrete colatitudinal index.  Then we compute a revised map, $B(\theta_k,t^\prime)$ by shifting each row in time such that $t^\prime = t - (\lambda - 40^\circ)/V_p$ where $\lambda = 90^\circ - \theta$ is the latitude in degrees and $V_p$ is a specified tracking speed.  

We then compute a correlation function between different rows of the shifted map as
\begin{equation}\label{cc1}
c(i,k; V_P) = \frac{\int B(\theta_i,t^\prime) B(\theta_k,t^\prime) dt^\prime}{\sqrt{\int B^2(\theta_i,t^\prime) dt^\prime \int B^2(\theta_k,t^\prime) dt'}}
\end{equation}
Here indices $i$ and $k$ span all latitudes between 40$^\circ$ and 70$^\circ$.  The time integration extends over three cycles, as shown in Fig.\ \ref{fig:all_bfly}.

\begin{figure}[!h]
\centering
\includegraphics[width=0.5\textwidth, angle=0]{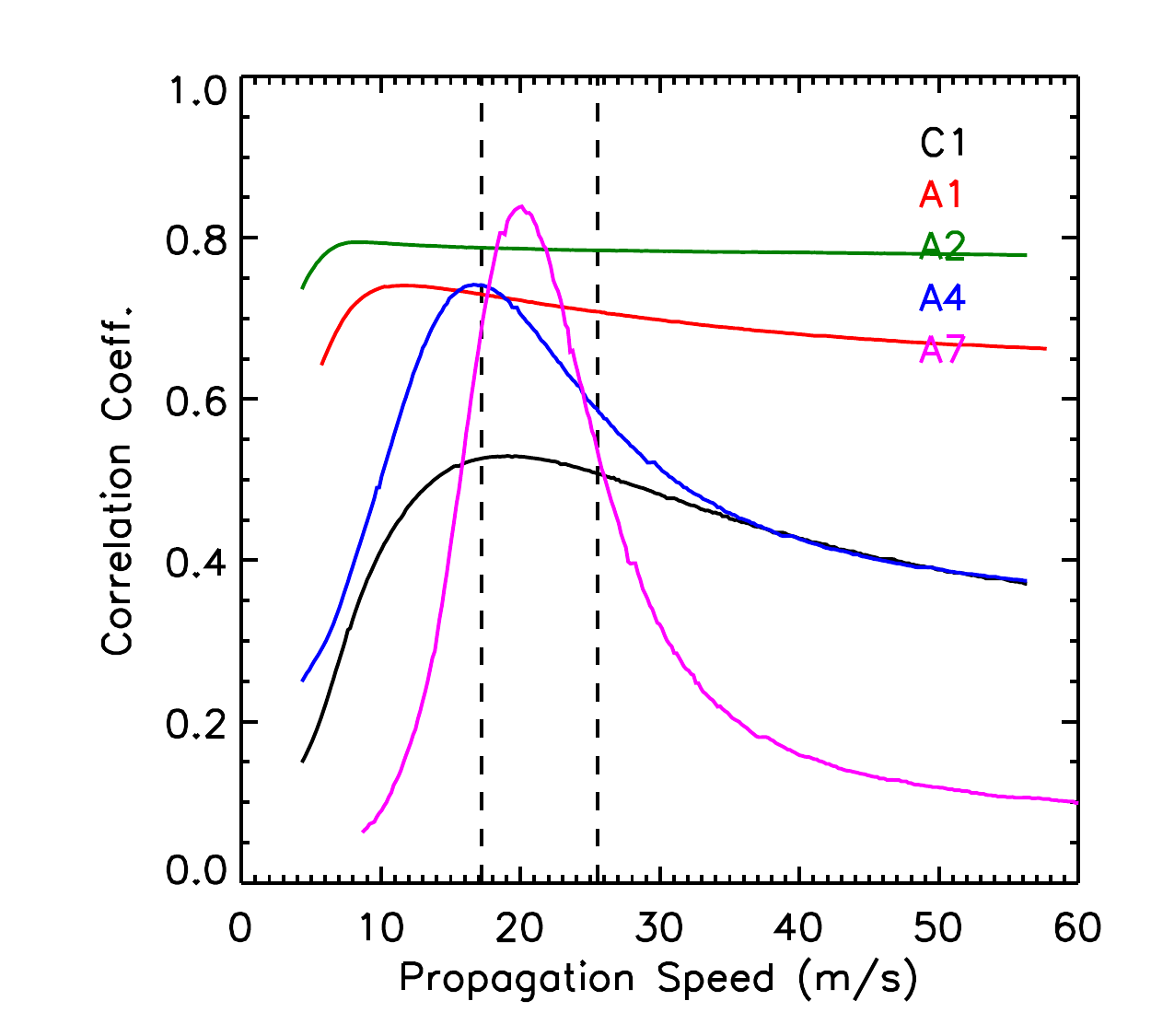}
\caption{The correlation coefficient $C$ versus migration speed $V_P$ is plotted for five different cases. The black solid line represents Case C1.  Red, green, blue and magenta lines represent Cases A1, A2, A4, and A7, as indicated.  The two dotted vertical lines show the minimum and maximum speed of the meridional circulation in the chosen latitude range of $40^\circ$--$70^\circ$.}\label{fig:migration}
\end{figure}

Then we sum the result of eq.\ (\ref{cc1}) over all $i$, $k$ pairs to obtain a single correlation coefficient for each value of $V_p$:
\begin{equation}
C(V_p) = \frac{1}{N} \sum_i \sum_{k>i} c(i,k;V_p)
\end{equation}
where N is the total number of pairs in the summation ($k>i$).  The final step is to plot $C(V_P)$ and choose the value of $V_p$ that maximizes the correlation as the characteristic flux migration speed for each case.  The results are listed in Table \ref{cases} and the correlation plots for several cases are compared in Fig.\ \ref{fig:migration}.

The meridional circulation is mainly responsible for the poleward transport of magnetic flux but convective transport can also play an important role. For those cases in which the convective transport is represented by a turbulent diffusion (A1-A8), one might expect that an increase in the diffusion coefficient $\eta_{top}$ might enhance the transport of flux toward the poles because both transport mechanisms (diffusion and MC) will work in concert.  However, we find the opposite; cases with higher diffusion have slower poleward migration speed (Table \ref{cases}).

The broad shape of the curve in Fig.\ \ref{fig:migration} for Case A1 reflects the general tendency for high levels of diffusion to smooth out the field.  The width of the streams is broad, so a range of $V_p$ values will give similar correlation coefficients.  However, there is a clear peak at about 11.6 m s$^{-1}$ that is substantially slower than the corresponding values of 19--20 m s$^{-1}$ for the less diffusive Cases A5, A7 and A8.  Furthermore, this migration speed is substantially slower than the meridional circulation speed (dashed lines), indicating that the presence of strong diffusion inhibits the poleward transport.

This result is akin to that discussed at the end of Section \ref{sec:nocon} with regard to the cycle period.  The higher value of diffusion implies a lower effective magnetic Reynolds number for flux transport near the surface.  For a velocity scale $U_{mc} \sim $ 20 m s$^{-1}$ at the surface and $L \sim \pi R / 2$, a value of $\eta_{top} = 3\times 10^{12}$ yields Rm $\sim 73$.  At this value of Rm, the magnetic field can slip through the plasma, violating Alfv\'en's theorem and rendering the poleward advection by the meridional flow less effective.  

\textit{The suppression of poleward transport by vertical turbulent diffusion is an effect that is not seen in 2D SFT models}.  It depends on the radial structure of the radial field so the radial dimension is needed to capture it.   While the surface meridional flow acts to advect flux poleward, the subsurface field lags behind, resisting this advection.  As the diffusion is decreased, the migration rate approaches the rate associated with advection by the meridional flow. 

This result was previously demonstrated for axisymmetric fields by \citet{guerr12} using 1D (latitude) and 2D (latitude, radius) advection-diffusion models for $\left<B_r\right>$.  They showed that the 1D surface transport model produced poleward flux migration at the rate of the meridional flow, regardless of the value of $\eta_{top}$.  However, in the 2D model, the poleward migration rate was systematically slower than the meridional flow of the plasma and the discrepancy increased with increasing $\eta_{top}$.  This is consistent with the parameter study of \citet{bauma04} who considered the effects of varying diffusion on 2D SFT models.  Close scrutiny of their Figure 6 reveals that the trailing polarity flux from mid-latitude BMRs begins to erode the pre-existing polar fields earlier in the cycle when the diffusion is large.  However, the timing of the polar field reversal is insensitive to $\eta_{top}$.  This is consistent with the idea that flux migrates poleward at a mean rate determined by the meridional flow but it also spreads, so that the leading edge of the flux migration reaches the poles sooner when the diffusion is large.

In some ways, solar observations appear to favor relatively high diffusion, as in Case A1.  The observed poleward migration speed of magnetic flux is about 10 m s$^{-1}$ \citep{howar81,topka82,hatha10b}, which is only about 60-70\% of the meridional flow speed \citep{ulric10}.  Furthermore, many 2D SFT models report optimal results when $\eta_{top}$ is on the order of 2--6 $\times 10^{12}$ cm$^2$ s$^{-1}$ \citep{wang89,jiang10b,Jiang_review15,lemer15}.  This latter value is justified by estimates of the diffusion rate from solar observations \citep{mosher77,topka82,schri96,abram11}.

The correlation curve in Figure \ref{fig:migration} for our convective Case C1 is suggestive of a high-diffusion case.  It is widely spread out and the peak value of 19.1 m s$^{-1}$ is slower than low-diffusive cases such as A7.  However, the correlation coefficient is significantly lower than in any of the diffusive cases and the characteristic migration speed is notably faster than comparable cases such as Case A1 (11.6 m s$^{-1}$). Interestingly, the cycle period for the Case C1 is higher than all of the diffusive cases including Case A1. How can these features be explained?

The higher poleward migration speed in Case C1 relative to Case A1 but slower migration speed of C1 compare to the low diffusivity A7 can be attributed to the vertical flux transport of the convective flows. The vertical flux transport in case of C1 is not efficient enough to be comparable with vertical transport in Case of A1 giving rise to faster migration speed but more efficient than the vertical transport of low diffusivity cases such as A7 suppressing the migration speed. As described earlier in this section, inefficient vertical transport of convective flows compare to the higher diffusion cases allows the surface flux to be advected poleward at a speed comparable to the meridional flow speed. Surprisingly, this is not consistent with Case C3, which has a lower value of $\eta_{top}$ than Case C1 and a slightly slower migration speed (Table \ref{cases}). The reason again may be related to the vertical transport of flux by the imposed convective flows fields. unlike the diffusion cases, the migration speed for the convective cases depend on the efficient vertical transport due to both diffusivity and convective flows. Therefore, when the background diffusivity less (e.g., Case C3), the vertical transport by convective flows may be higher and the resultant vertical transport is much efficient than the case C1 with slightly higher diffusivity (see Table~\ref{cases}).

The broad profile of the correlation curve for Case C1 in Fig.\ \ref{fig:migration} reflects the efficient horizontal transport.  As in Cases A1 and A2, the poleward streams are broad and diffuse, giving similar correlation coefficients for a range of tracking speeds.  The low value of the correlation coefficient ($< 0.55$) reflects the presence of residual mixed polarity from both leading and trailing spots, as we will discuss further in Section \ref{sec:energetics}.  We also address the cycle period of Case C1 in Section \ref{sec:energetics}, which is linked to the issue of mixed polarity and dynamo efficiency.

\subsection{Dynamo Efficiency}\label{sec:energetics}
As in any hydromagnetic dynamo, the field strengths achieved in our simulations are determined by a balance between magnetic field generation, nonlinear saturation, and ohmic diffusion.  In our model, the ohmic diffusion operates through the turbulent diffusivity $\eta$ which is varied by altering its value in the upper CZ, $\eta_{top}$.  All simulations have the same nonlinear saturation mechanism, as expressed in eq.\ (\ref{flux}), the same mean flows, and the same parameters for the flux emergence algorithm, SpotMaker, as discussed in Section \ref{sec:STABLE}.  So, one would expect simulations with lower diffusion to achieve higher field strengths.  And, this is indeed the trend seen for Cases A1-A8 in Table \ref{cases}.

Case C1 also falls within this general picture.  The mean field strengths achieved here are similar to Case A4 (Table \ref{cases}), which has a comparable value of $\eta_{top}$.  Thus, convection can transport fields, as demonstrated by the diffuse appearance of Fig.\ \ref{fig:all_bfly}\textit{a} compared to Fig.\ \ref{fig:all_bfly}\textit{d} and the broad profile of the correlation curve in Fig.\ \ref{fig:migration}, but it does not greatly enhance the ohmic dissipation.  Although Case C1 resembles Case A1 in other respects (see Sections \ref{sec:con} and \ref{sec:migration}), it's efficiency (as measured by the strength of the mean fields) is more comparable to Case A4.

The tendency for the convection to disperse fields without dissipating them is also reflected in the lower magnetic energy in the non-axisymmetric field component in Case C1 (2.87 kG) relative to Case A4 (4.64 kG).  This aspect of turbulent diffusion was emphasized in a series of papers by \citet{Piddington75, Piddington76, Piddington81}.   He argued that turbulent diffusion may not be good representation of merging and cancellation of fields, possibly over-estimating the cancellation rate by as much as 0.3 $R_m$ \citep{Piddington81}.  

In the context of our simulations, this implies that both polarities present in a particular BMR will be dispersed by convective motions, with little cancellation. The mixed polarity in Case C1 is particularly apparent at the poles.  As mentioned in Section \ref{sec:transport}, the polar regions are surrounded by a region of opposite polarity formed from leading flux that has been advected poleward along with the trailing flux.  The low value of $\eta_{top}$ implies a small ohmic dissipation scale; flux elements must come into close proximity in order to cancel.  So, the polar field strength in Case C1 (Fig.\ \ref{fig:bfly_con}\textit{b}) is about a factor of three higher than in Case A1 (Fig.\ \ref{fig:bfly1}\textit{b}; as is the mean poloidal field--see Table \ref{cases}), and this in turn promotes the generation of strong toroidal fields.

However, when the flux in Case C1 is concentrated at the poles, it does eventually cancel. This may explain why the cycle period in Case C1 (17.4 yrs) is higher than in Case A4 (15.1 yrs) even though they have similar values of $\eta_{top}$. Even the cycle period is higher than the case A8 (lowest $\eta_{top}$ used in our simulation). The concentrated mixed polarity fields near the polar region takes a longer time for polar field to reverse. As the migration speed of the horizontal flux is faster and ohmic dissipation due convective flows is not efficient as turbulent diffusion, the polar regions are surrounded by residual leading polarity flux along with the trailing polarity flux. The trailing polarity flux get canceled out due to residual leading polarity flux and effectively less trailing polarity fluxes are available to reverse the existing polar field. Therefore for proper polar fields reversal, a large accumulation of trailing polarity flux is necessary which needs longer time.

In summary, a turbulent diffusion coefficient of $\sim 3 \times 10^{12}$ cm$^2$ s$^{-1}$ as in Case A1 adequately captures the surface flux transport in Case C1 but it does not adequately capture the dissipation of magnetic energy. Approximating convective transport with a turbulent diffusion will likely have an adverse effect on the dynamo efficiency, producing artificially weak mean fields and shorter cycles.

\section{Summary and Conclusion}\label{sec:summary}

Turbulent transport of vertical magnetic flux by near-surface convective motions is an essential component of BL dynamo models.  In particular, it plays an important role in generating the polar fields that are the seed for the next cycle \citep{CCJ07}.  This process is captured with high fidelity by the AFT Surface Flux Transport (SFT) model, which simulates the advection of vertical magnetic flux by a horizontal flow field that is based on the observed convective power spectrum in the solar photosphere \citep{upton14a,upton14b,ugart15}.  In this paper we have taken a substantial step forward in the unification of BL dynamo models and SFT models by presenting the first BL dynamo model that incorporates a realistic photospheric convection spectrum.  

Our 3D BL dynamo model, STABLE, uses the same surface flow field as AFT, though unlike AFT, ours is independent of time.  The generalization to an evolving field as used by AFT is straightforward and will be implemented in future work.  In our initial implementation, we extrapolated this surface flow field downward based on mass conservation, producing a vigorous 3D convective flow that permeated the upper portion of the convection zone  (Sec.\ \ref{sec:fr}).  However, we quickly found that this caused problems for our kinematic framework.  In particular, we found that this 3D convective flow field is an efficient small-scale dynamo, producing a chaotic field component that grew exponentially, overwhelming the cyclic field component (Sec.\ \ref{sec:ssd}).  Note that our implemented convective flow is non-helical, so there is no turbulent $\alpha$ effect, but it is capable of generating small scale fields by stretching and shearing of the magnetic fields near upper convection zone. One could in principle suppress this by increasing $\eta_{top}$ until the effective value of the magnetic Reynolds number is subcritical to dynamo action.  However, achieving $R_m \sim 50$ would require a value of $\eta_{top}$ of about $1.7 \times 10^{12}$ cm$^2$ s$^{-1}$, which is comparable to the estimate of convective transport from photospheric observations (see Sec.\ \ref{sec:intro}).  Since our objective is to {\em replace} turbulent diffusion with a more realistic depiction of convective transport, such a large value of $\eta_{top}$ is undesirable.

Instead we chose to suppress small-scale dynamo action by only applying the convective flow field to the radial component of the magnetic field.  In addition, we kept a significant background diffusion of $\eta_{top} = 5 \times 10^{11}$ cm$^2$ s$^{-1}$ (see black dotted line in Fig.~\ref{fig:eta}) in the upper layers of the convection zone ($r > 0.9R$) as well as an enhanced hyperdiffusion. We found that we could achieve regular magnetic cycles with a lower background diffusion ($\eta_{top} = 5 \times 10^{10}$ cm$^2$ s$^{-1}$) but the dynamo switched to a quadrupolar parity.

Through this approach we were able to achieve viable, cyclic, solar-like dynamo models.  The general appearance and behavior of these models is similar to non-convective cases with a surface diffusion $\eta_{top} \sim 3 \times 10^{12}$ cm$^2$ s$^{-1}$ (Case A1; Secs.\ \ref{sec:transport}, \ref{sec:migration}).  This is demonstrated in particular by the flat profile of the correlation function for Case C1 shown in Figure \ref{fig:migration}.  This value of $\eta_{top}$ is comparable to the range of 2--6 $\times 10^{12}$ cm$^2$ s$^{-1}$ that is estimated from solar observations and that is often used in SFT models \citep{mosher77,topka82,wang89,schri96,jiang10b,abram11,Jiang_review15,lemer15}.

Treating the BL mechanism with explicit convective transport (Case C1) gives us more realistic surface flux transport in comparison to the case with an equivalent surface diffusion (Case A1). The BMRs that emerge at active latitudes are fragmented and dispersed by the convective flows, with less flux cancellation.  However, Case C1 does exhibit a band of opposite polarity surrounding the polar cap that is not seen is solar observations, at least not to this extent (Fig.\ \ref{fig:bfly_con}\textit{a}).  This mixed polarity in the polar region is not seen in the case with turbulent diffusion. We attribute its existence to residual leading flux that is advected poleward along with the trailing flux.  

In our models, the speed at which residual poloidal flux from BMRs migrates to the poles is determined mainly by the vertical diffusion, $\eta_{top}$.  For small values of the $\eta_{top}$ ($\lesssim 8 \times 10^{11}$ cm s$^{-1}$), this poleward migration speed approaches the meridional flow speed.  However, for high vertical diffusion (efficient mixing), the migration speed is slower, as surface flux transport is impeded by the subsurface field that is being ``dragged along''.

Several aspects of Case C1 highlight the limitations of parameterizing convective transport as a turbulent diffusion.  First, we found that the direction and amplitude of the turbulent emf was not well correlated with the local gradient of $\left<B_r\right>$ (Sec.\ \ref{sec:emf}).  In order words, the transport of magnetic flux by the convective motions was not in general down-gradient, as would be expected from a diffusive flux.  However, we attribute this to the tendency of convective flows to advect vertical flux into ``downflow lanes'', or more precisely, regions of horizontal convergence, before dispersing it.  So, this cannot rule out the possibility that the convective transport is diffusive on scales much larger than that of the convective motions ($\gtrsim$ 34 Mm), as expected from mean-field theory.

A second limitation of the turbulent diffusion parameterization is an over-estimate of the ohmic dissipation (Sec.\ \ref{sec:energetics}).  The mean fields in the convective Case C1 are about a factor of three stronger than in Case A1, which has $\eta_{top} = 3 \times 10^{12}$ cm$^2$ s$^{-1}$.  This is particularly true for the peak polar fields, which are an important factor in determining the strength of the following cycle.  This over-estimate of the ohmic dissipation, or alternatively, an under-estimate of the dynamo efficiency, cannot be addressed with traditional SFT models; since these are not dynamo models, the field strengths are not regulated by the same interplay between field generation, nonlinear saturation, and ohmic dissipation.  Thus, it is a new result that has not been identified in previous studies.

In conclusion, the use of explicit convective motions is a promising way to improve the fidelity of BL dynamo models that have the capability to model 3D flows.  However, the main challenge to producing viable models of the solar cycle with this approach is to properly handle the small-scale dynamo action that will likely ensue.  This may ultimately require a more consistent MHD formulation that takes into account flow suppression by Lorentz force feedbacks.


We thank David Hathaway and Lisa Upton for providing the surface convective velocity fields and the images in Figs.\ \ref{fig:conhath}\textit{a},\textit{b}.  We also thank Arnab Rai Choudhuri, Lisa Upton and Bidya Binay Karak for many informative discussions throughout the course of this work. GH thanks CSIR, India for financial support. Computational resources were provided by NCAR and NASA's High-End Computing (HEC) program.  The visualizations in Figures \ref{fig:viz1}\textit{c} and Fig.\ \ref{fig:viz2} were made using NCAR's Vapor software package: http://www.vapor.ncar.edu.  NCAR is sponsored by the National Science Foundation.

\bibliography{myref,Markrefs}
\end{document}